\documentclass[twocolumn,showpacs,preprintnumbers,amsmath,amssymb]{revtex4}

\usepackage [dvips]{graphicx}
\usepackage{dcolumn}
\usepackage{bm}

\begin{document}

\preprint{APS/123-QED}

\title{Spectral Analysis of the Growth of Translational Superfluid Flow in a Bose Liquid }

\author{Shun-ichiro  Koh}
 \affiliation  {Physics Division, Faculty of Education, Kochi University, 
 Akebono-cho, 2-5-1, Kochi, 780, Japan.}

 \email{koh@kochi-u.ac.jp}

\date{\today}

\begin{abstract}

Spectral analysis of the translational superfluid flow of a Bose liquid is attempted.
When cooling  a dissipative flow of  liquid helium 4 through a capillary at $T>T_{\lambda}$,
a superfluid flow  abruptly appears at $T_{\lambda}$.  By a thought 
experiment in which the pressure difference between two ends of a capillary slowly oscillates, 
the spectrum of fluidity  (the reciprocal of kinematic viscosity) just above $T_{\lambda}$
 ($2.17 K<T<2.18 K$) is examined using a sum rule that incorporates the 
 result of linear response theory and fluid  mechanics. 
According to  Feynman's picture on the many-body wave function of bosons,  
as the condensate grows in the flow, the transverse  motion in the flow becomes suppressed
by the emergence of a gap, originating from the Bose statistics, in the excitation spectrum.
  This statistical  gap induces a continuous change in the fluidity spectrum just above $T_{\lambda}$,  
which demonstrates the growth  of a superfluid flow.   For this 
  mechanism, the size distribution  of the coherent wave function plays an important role. 
  As a byproduct, a new interpretation of the critical velocity is proposed.
\end{abstract}


\maketitle

\section {Introduction}

\begin{figure}
		\begin{center}
\includegraphics [scale=0.5]{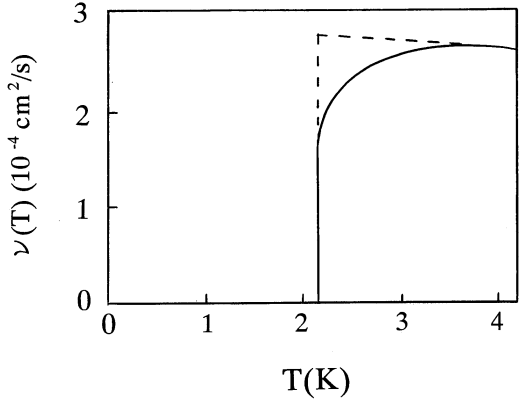}
\caption{\label{fig:epsart} 
 The coefficient of kinematic viscosity $\nu (T)$ (solid curve) of  the capillary 
 flow of liquid helium 4.  [The dashed line illustrates $\nu (T)$ when the 
 gradual decrease in $\nu (T)$ does not occur.]}
   \label{fig.005¥}
     	\end{center}
\end{figure}

Superfluidity has been a mysterious phenomenon 
since the discovery of a superfluid flow in liquid helium 4 in 1938
\cite{kap} \cite{land} \cite{lon}. It is classified into 
a translational superfluid flow and  rotational superfluid flow. When 
 the  temperature is decreased to $T_{\lambda}$, the former appears in the flow 
 which is until then dissipative at $T>T_{\lambda}$ as in a flow in a capillary, 
and the latter appears in the flow which is until then nondissipative (in principle)   
at $T>T_{\lambda}$ as in a flow in a rotating  bucket. In comparison 
with the rotational superfluid flow,
questions  still remain  to be answered in the translational 
superfluid flow,  because it is more directly  concerned with  
 the irreversible processes in liquids \cite{rec}.
 When cooling  liquid helium 4 flowing in a capillary, the coefficient of kinematic 
viscosity $\nu (T)$ gradually decreases in the range of $T_{\lambda}<T<3.7$ K,
finally reaching zero at  $T_{\lambda}$, as shown in 
Fig. \ref{fig.005¥} \cite{men}\cite{zin}\cite{oth}.
In the range of $2.17$ K $<T<2.18$ K, superfluid flow  abruptly grows
within a narrow temperature region of less than $0.01$ K.
Behind this seemingly  abrupt decrease in  viscosity, we can see a 
remarkable  phenomenon: When the external parameter falls below the critical value, the complex 
 process transforms to a simple one  by the formation of order (the inverse of 
 bifurcation \cite {inv}). In this paper, we  derive $\rho _s\mbox{\boldmath $v$}_s$ in the 
 two-fluid mechanics from the normal flow just above $T_{\lambda}$, 
 and consider  the transition of a laminar viscous flow to a 
 translational superfluid flow of a Bose liquid from viewpoint of spectrum analysis.

(a) In the normal phase, the flow in a capillary is described as Poiseuille's flow in fluid mechanics. 
When a pressure difference $P$ between two ends of a capillary 
(circular cross section with radius $d$, and length $L$) is applied to a normal liquid 
with density $\rho$ and coefficient of shear viscosity $\eta$, 
the Navier--Stokes equation with Stokes's approximation, 
\begin{equation}
 \rho\frac{\partial\mbox{\boldmath $v$}}{\partial t¥}¥= \eta¥ 
 \nabla^2\mbox{\boldmath $v$}- \nabla P ,
          \label{eq:002¥}
\end{equation}¥
gives the fluid velocity of a steady flow as a function of radial 
distance $r$ from the axis of capillary as follows:
\begin{equation}
 \mbox{\boldmath $v$}(r)= \frac{d^2-r^2}{4\eta¥}¥\frac{\mbox{\boldmath $P$}}{L¥} .
          \label{eq:005¥}
\end{equation}¥
This formula is a dissipative-type relation $v \propto P$, similar to 
a terminal velocity which a raindrop falling from the atmosphere 
reaches  before falling to the ground. 
Let us  view this system on a reference frame  moving with velocity
$\mbox{\boldmath $v$}$.  To study the change from a normal flow to a
superfluid flow, we consider the longitudinal motion and the transverse motion 
  in normal flow, and decompose them  into excitations
 with different momenta and frequencies. 
The infinite-wavelength limit ($q\rightarrow 0$)
 of the longitudinal excitation  is equivalent to the uniform 
 translation of a liquid. Hence, at this limit, excitations with small 
 momenta and frequencies are smoothly 
 transformed to a translational superfluid flow. The growth of a superfluid flow is a 
process in which other excitations with finite $q$, 
contained in the viscous flow at $T>T_{\lambda}$, 
converge to the superfluid flow with $q=0$. Although the 
superfluid flow is a steady flow, the spectrum analysis of the viscous flow 
 near $T_{\lambda}$ gives us important information on the growth of 
superfluid flow.

\begin{figure}
		\begin{center}
\includegraphics [scale=0.55]{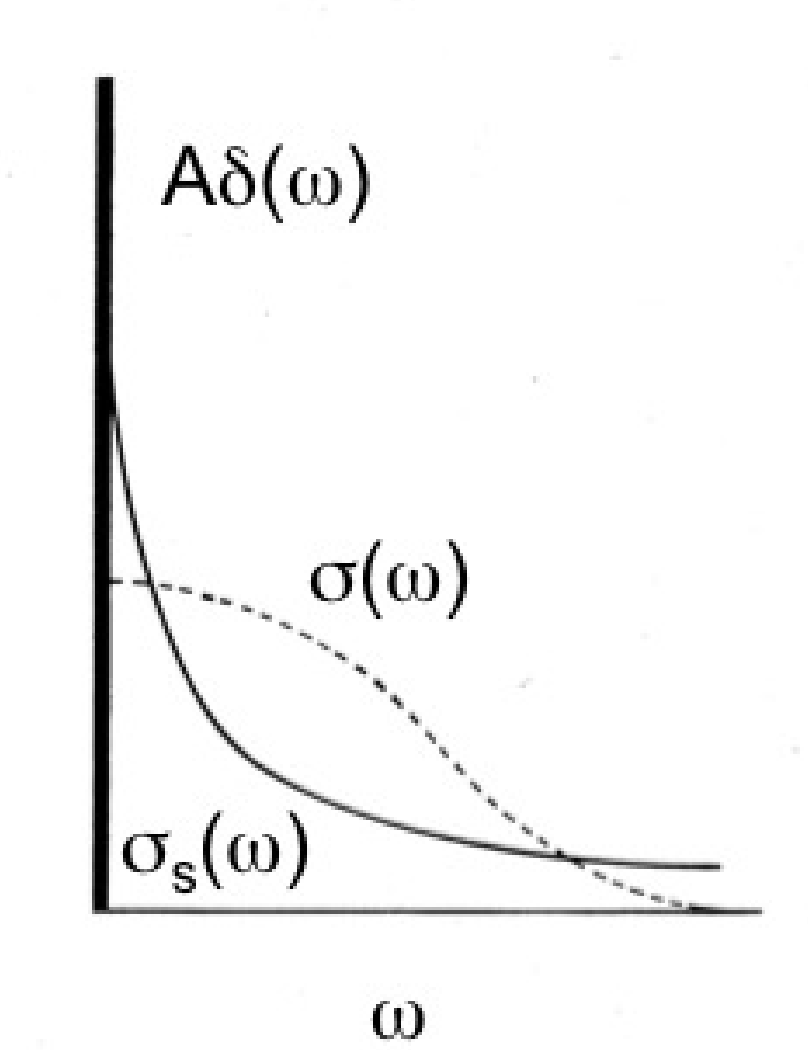}
\caption{\label{fig:epsart} Schematic view of the fluidity spectrum
$\sigma _s(\omega)¥+A\delta (\omega)$ characterizing a superfluid flow. 
The dotted curve represents  $\sigma (\omega)$ for a normal liquid in 
Eq.(\ref{eq:293¥}). $\sigma(0)$ is inversely proportional to $\nu$ in Fig.\ref{fig.005¥}.}
\label{fig.015¥}
     	\end{center}
\end{figure}

Along this line, we consider a thought experiment in which 
a slowly  oscillating pressure difference is applied  to a
 fluid as $P(\omega)\exp (-i\omega t)$, which is an extension of 
 the oscillatory Poiseuille's flow experiment \cite {tur}. 
In the case of a normal liquid, an oscillating mass-flux density $j(\omega)=\rho v(\omega)$ 
appears along the axis of the capillary in Eq.(\ref{eq:005¥})  \cite {axis},
\begin{equation}
 \mbox{\boldmath $j$}(\omega)=\sigma (\omega)¥¥d^2 
              \frac{\mbox{\boldmath $P$}(\omega)}{L¥} ,
          \label{eq:01¥}
\end{equation}¥
where $\sigma(\omega)=\rho/(4\eta)=1/(4\nu)$ is the frequency spectrum 
of fluidity, the dimension of which is $(length)^{-2}(second)$.
[From now, we simply call $j(\omega)$ a flow, and call 
$\sigma (\omega)¥$ a {\it  fluidity  spectrum\/} of a liquid.] 
This $\sigma(\omega)$ of the normal liquid is calculated in fluid 
mechanics as follows.
Under the oscillating pressure gradient $ P(\omega) \exp (-i\omega 
t)/L$, Eq.(\ref{eq:002¥}) is written in cylindrical polar coordinate as 
 \begin{equation}
 \frac{\partial v}{\partial t¥}¥=\nu \left( \frac{\partial }{\partial r^2¥}
                     + \frac{\partial }{r\partial r¥}\right)v
                       + \frac{ P(\omega) \exp (-i\omega t)}{\rho¥L}¥,
							  	\label{eq:2905¥}
\end{equation}¥
where $\nu=\eta /\rho $. The velocity $v(r,t)$ at a distance $r$ from 
the axis has the form
\begin{equation}
 v(r,t)¥=\frac{ P(\omega)\exp (-i\omega t)}{-i\omega \rho¥L}+\delta v(r,t)¥,
                    	\label{eq:29071¥}
\end{equation}¥
where $\delta v(r,t)$ satisfies 
\begin{equation}
 \frac{\partial \delta v(r,t)}{\partial t¥}¥=\nu \left( \frac{\partial}{\partial r^2¥}
                  + \frac{\partial }{r\partial r¥}\right) \delta v(r,t)¥.
					    	\label{eq:291¥}
\end{equation}¥
The solution $\delta v(r,t)$ of Eq.(\ref{eq:291¥}) is written in 
terms of the Bessel function as $J_0(i\lambda _0r)\exp (-i\omega t)$
with $\lambda _0=(1-i)\sqrt {\omega /(2\nu)}$ \cite {sex}. Using this 
$J_0(i\lambda _0r)\exp (-i\omega t)$ as $\delta v(r,t)$, 
and setting $v(d,t)=0$  in Eq.(\ref{eq:29071¥}) as a boundary condition, we obtain
\begin{equation}
 v(r,t)=\frac{ P(\omega)\exp (-i\omega t)}{-i\omega \rho¥L¥}¥
                       \left(1-\frac{J_0(i\lambda _0r)}{J_0(i\lambda _0d)¥}\right)¥.
					    	\label{eq:2915¥}
\end{equation}¥
With this $v(r,t)$ at $r=0$ used in $\rho v(0,t)=\sigma  (\omega )d^2  P(\omega)\exp (-i\omega t)/L$ 
of  Eq.(\ref{eq:01¥}), the fluidity spectrum  $\sigma (\omega )$ in the 
oscillatory Poiseuille's flow is given by  
 \begin{equation}
 \sigma(\omega) =\frac{-1}{\omega d^2¥}Im\left(1-\frac{1}{J_0\left(id[1-i]\displaystyle {\sqrt 
 {\frac{\rho}{2\eta ¥}¥\omega}}\right)¥¥}\right)¥,
          \label{eq:293¥}
\end{equation}¥
which is schematically shown as a dotted curve in Fig.\ref{fig.015¥}. 
The fluidity of a flow through a capillary  
gradually falls with the increase in frequency of the  pressure oscillation. 
 Expanded in powers of $\omega$,  $\sigma (\omega)$ in 
 Eq.(\ref{eq:293¥}) has the following form
\begin{equation}
   \sigma (\omega)=\frac{\rho }{4\eta_n¥}
    \left[1-\frac{d^4}{576¥}¥\left(\frac{\omega}{\eta_n/\rho¥}¥\right)^2¥+ \cdots\right]¥.
                  	\label{eq:361¥}
\end{equation}¥
This $\sigma(\omega)$  does not substantially fall 
near $\omega =0$ [in this case,  $\eta_n /(\rho d^2)\simeq 2\times 10^{-2}$ [1/s]],
implying that the viscous flow is robust  when the applied pressure slowly oscillates. 
In this paper, we  address a question of how  the growth of a superfluid flow 
changes this feature.

(b) In the superfluid phase, owing to the growth of a condensate, the 
$\rho v \propto P$ -type relation in Eq.(\ref{eq:01¥}) changes to a nondissipative 
Newtonian law $\rho a\propto P$, where $a$ is acceleration.
 This change  implies that once pressure is applied to this liquid at 
 the beginning, a nonzero flow ($j\neq 0$) continues to exist even  
after the pressure difference is removed ($P=0$).   
 Such a phenomenon can be described by Eq.(\ref{eq:01¥}) only when $\sigma (\omega)$ 
 shows  divergence at $\omega =0$. [The abrupt increase in $\sigma$ 
 implies the vertical fall of $\nu =1/(4\sigma)$ as in Fig.\ref{fig.005¥}.] 
 This divergence at $\omega =0$ is not an isolated phenomenon in the frequency.  
  Since the values of $\sigma  (\omega)¥$  at finite $\omega$ and at $\omega =0$  
  represent the same growth process at different frequencies, 
 the most probable form of the fluidity spectrum in a superfluid flow is illustrated
in Fig.\ref{fig.015¥}, which is given by
\begin{equation}
 \mbox{\boldmath $j$}(\omega)=\left[\sigma _s(\omega)+A(T)\delta(\omega)\right]¥¥d^2 
 \frac{\mbox{\boldmath $P$}(\omega)}{L¥},
	    \label{eq:02¥}
\end{equation}¥
where $A(T)$ is a constant to be determined.  $\sigma _s(\omega)$ includes  
information on the precursory form of a superfluid flow of mesoscopic size.  An interesting point is 
whether this $A(T)$ is different from the thermodynamic rate 
$\rho _s(T)/\rho$.  This $\sigma _s(\omega)+A(T)\delta(\omega)$ is a sign of 
a translational superfluid flow emerging in the dissipative process.  
As an early study along this line, we studied the frequency-expansion 
of $\sigma _s(\omega)+A(T)\delta(\omega)$ in the vicinity of $T_{\lambda}$  
\cite {ear}. In the present paper, by considering the effect of Bose 
statistics on the flowing coherent many-body wave function more 
concretely, we will derive  a precise form of the fluidity spectrum 
in all frequencies, and incorporate it into a sum rule.

To study the change from Eq.(\ref{eq:01¥}) to Eq.(\ref{eq:02¥}), it 
is useful to focus on a relation holding both at $T>T_{\lambda}$ and 
$T\leq T_{\lambda}$.  The fluidity spectrum  must satisfy a  sum 
rule.  To illustrate its physical meaning, let us consider the 
following equation of motion under pressure with a relaxation time $\tau¥$:
\begin{equation}
  m\frac{dv}{dt¥}¥=Pd^2-\frac{m}{\tau¥}¥v .
	   	\label{eq:297¥}
\end{equation}¥
Under the  oscillating pressure $P\exp (-i\omega t)$, the oscillating velocity is given by
\begin{equation}
  v(\omega)=\frac{Pd^2}{m¥}¥\left(¥\frac{\tau^{-1}}{\omega^2+\tau^{-2}¥}¥
                                 +i\frac{\omega}{\omega^2+\tau^{-2}¥}\right).
	   	\label{eq:2974¥}
\end{equation}¥
Comparing it to $j(\omega)=\rho v(\omega)=\sigma (\omega)d^2P/L$, we obtain
\begin{equation}
 \sigma (\omega)=\frac{\rho L}{m¥}¥\left(¥\frac{\tau^{-1}}{\omega^2+\tau^{-2}¥}¥
                                 +i\frac{\omega}{\omega^2+\tau^{-2}¥}\right)  .
	   	\label{eq:2976¥}
\end{equation}¥
Hence, we obtain the sum rule
\begin{equation}
   \int_{-\infty¥}^{\infty¥}Re\sigma(\omega)d\omega = \pi\frac{\rho L}{m¥}¥.
	   	\label{eq:2978¥}
\end{equation}¥
In this derivation, $\tau$ is necessary, but it disappears in 
the final result. The right-hand side of Eq.(\ref{eq:2978¥}) includes 
only parameters in setting experiments, which is a feature of general 
relations (see Appendix A). Hence, 
 the change in the fluidity spectrum $\sigma(\omega)$ from 
Eq.(\ref{eq:01¥}) to Eq.(\ref{eq:02¥}) around $T_{\lambda}$ must satisfy 
\begin{equation}
 \int_{-\infty¥}^{\infty¥}\sigma(\omega)d\omega
    =\int_{-\infty}^{\infty¥}[\sigma _s(\omega)¥+A(T)\delta(\omega)] d\omega .
			               \label{eq:2983¥}
\end{equation}¥
Substituting the concrete formula of the fluidity spectrum into both sides, we 
can examine  how the superfluid flow grows out of the viscous flow.

(c) Behind the above change in $\sigma (\omega)$, 
some physical  process  must exist.  A superfluid flow 
withstands  thermal dissipation, which implies that there must be a 
gap in the excitation-energy spectrum. (Do not 
confuse this gap with the roton gap in the phonon spectrum of liquid helium 4.) 
In two-fluid mechanics, Landau  assumed the 
irrotationality constraint $rot \mbox{\boldmath $v$}_s =0$ on the 
superfluid velocity, which implies that the low-energy excitation of a transverse 
nature is suppressed in a superfluid. Hence, such a gap must exist in the 
spectrum of transverse excitation. 
In this respect, a suggestive consideration was given by Feynman in 
the 1950s \cite {fey1}. Despite the dissipation of energy in liquids, 
the longitudinal spectrum (that is, acoustic phonon) was clearly 
observed by the inelastic neutron diffraction in the 
superfluid phase of liquid helium 4. 
The central problem is explaining  why no states other 
than the acoustic phonon can have low energies. Feynman considered that  
an energy gap owing to Bose statistics clearly separates the spectrum  
of the acoustic phonon from that of other excitations. 
Consider a free particle inside a box of size $L_0$. The lowest 
eigenvalue of the Laplacian  is of the order of $\sim 1/L_0^2$. This is 
realized by an excitation with the longest possible wavelength $\lambda 
\sim L_0$. When $\lambda $ is finite, the spectrum $\epsilon (q)$ of 
such a system  has a gap at $q = 0$. 
Consider the displacement of particles in the condensate. When the
permutation symmetry of Bose statistics holds, multiple permutation, 
operated on the distribution of particles  after displacement, results in a 
final distribution which is apart from the initial displacement only by a 
small distance similar to the interatomic distance $r_0$ in 
liquid. Hence, a large $L_0$ is not possible, and  such a $L_0$ is 
reduced to the interatomic distance $r_0$ in a liquid.
The kinetic energy,  determined by the square of the spatial 
derivative of the wave function $\psi$, cannot have an excitation 
energy lower than a limit as follows:
\begin{equation}
  \frac{\hbar ^2}{2m¥}¥(\nabla\psi )^2 \geq  \frac{\hbar 
  ^2}{2m¥}¥\frac{(\delta\psi)^2}{r_0^2¥}¥ .
	     	\label{eq:053¥}
\end{equation}¥
In liquid helium 4, where the interatomic distance  
 is $r_0\simeq 5\times 10^{-11}$ [m] and the mass of a helium 4 atom is 
 $m=6.7\times 10^{-27}$ [kg] \cite {don}, we obtain 
$\epsilon _0\equiv(\hbar ^2/2m)¥(1/r_0^2)\simeq 3.0\times 10^{-22}¥$ [J]. 
This $\epsilon _0$ is much larger than $\hbar \omega$ realized in the 
hydrodynamic experiments, and therefore 
this  $\epsilon _0$ completely suppresses such a transverse  excitation  
in a Bose liquid. Since this gap is due to Bose statistics, not to the particle interaction, 
we will call this $\epsilon _0$ a statistical gap.

This explanation is also applicable to the nonequilibrium steady process. 
The problem is whether  such a permutation is possible on a macroscopic scale in the flow.
In the Bose system, permutation is always possible locally between 
adjacent particles. However,  whether macroscopic permutation between distant particles is 
possible    depends on the type of flow pattern. 
Only when  the displacement vectors are  transverse with respect to 
the flow direction, permutation on the macroscopic scale is possible. 
Hence, the point to be considered is whether the transverse excitation occurs within or 
beyond the coherent wave function. 
Near $T_{\lambda}$,  various sizes of coherent wave functions appear in the Bose 
liquid. Using their size distribution in this 
paper, we  will clarify the mechanism by which the statistical gap in the 
transverse motion  changes the  fluidity spectrum from $\sigma (\omega)$
to $\sigma _s(\omega)+A(T)\delta(\omega)$, and 
explain the growth process of a superfluid flow in the range of $2.17$ K $<T<2.18$ K. 

Phenomena near $T_{\lambda}$ are affected by fluctuation. However, 
for the fluctuation in  nonequilibrium 
steady states, an example of which is a flowing liquid  near $T_{\lambda}$,  
 a convincing theoretical description has yet to be found \cite {eva}.
Hence, we ignore such a fluctuation in this paper, and regard  
results obtained in this paper as a first step to the complete statistical theory.

This paper is organized as follows. In Sec.2, we explain the nature of 
the statistical gap appearing in the transverse motion in the flow, and 
introduce the size distribution of the coherent many-body wave function.
In Sec.3, we formulate  Poiseulle's flow by rewriting Eq.(\ref{eq:01¥}) as a linear-response 
relation, then formulating  the two-fluid model in terms of the fluidity spectrum. 
In Sec.4, we derive the change in the fluidity  spectrum from $\sigma (\omega)$ to 
$\sigma _s(\omega)+A\delta(\omega)$ in Fig.\ref{fig.015¥}
by combining the linear response theory and fluid mechanics in the sum rule.  
 In Sec.5, we perform a model calculation using the result of Sec.4.
 In view of the concrete form of $\sigma _s(\omega)$ in Sec.5.D, a new interpretation of 
 the critical  velocity of a superfluid flow is proposed. In Sec.6, we discuss remaining problems.

\section { Physical View of Translational Superfluid Flow}

\subsection  {Longitudinal motion and transverse motion in the flow}

 \begin{figure}
		\begin{center}
\includegraphics [scale=0.5]{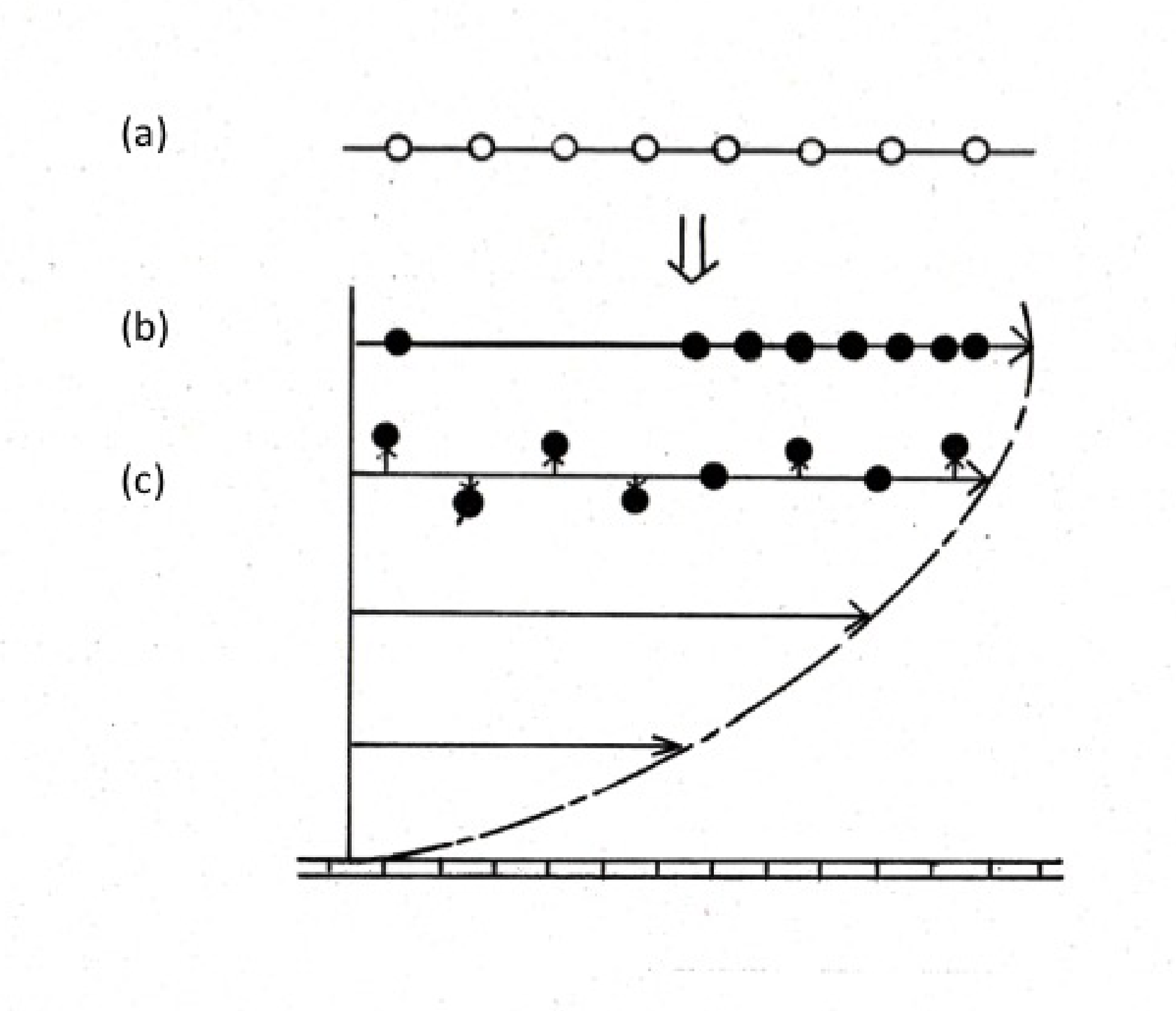}
\caption{\label{fig:epsart}   Schematic view of the microscopic relative 
displacement of particles in a normal capillary flow when one views 
them on a moving frame: (b) longitudinal and (c) transverse 
displacements  from (a) the initial uniform distribution. 
 The real image  is a mixture of these two types of displacements, 
the lengths of which are exaggerated for clarity. 
The macroscopic one-directional velocity $\mbox{\boldmath $v$}(r)$ of Poiseuille's flow 
[Eq.(\ref{eq:005¥})] is illustrated for reference. }
\label{fig.1¥}
     	\end{center}
\end{figure} 

(a) Figure \ref{fig.1¥} schematically illustrates particles in a normal liquid flowing 
through a capillary. When one views them on a moving frame from 
left to right, one notes two types of microscopic relative displacements of particles. 
 Let us imagine a relative position $(\mbox{\boldmath $x$}_1,\cdots , \mbox{\boldmath $x$}_s)$ 
 of $s$ particles on a streamline, which changes  
to $[\mbox{\boldmath $x$}_1+\delta\mbox{\boldmath $R$}(\mbox{\boldmath $x$}_1) ,\cdots , 
  \mbox{\boldmath $x$}_s+\delta\mbox{\boldmath $R$}(\mbox{\boldmath $x$}_s)]$.
 We can  classify the relative displacements $\delta\mbox{\boldmath $R$}(\mbox{\boldmath $x$}_i) $ 
into  longitudinal $\delta\mbox{\boldmath $R$}^l$ and transverse 
$\delta\mbox{\boldmath $R$}^t$ displacements such that 
$\delta\mbox{\boldmath $R$}^l\parallel \mbox{\boldmath $q$}$ and 
$\delta\mbox{\boldmath $R$}^t \perp \mbox{\boldmath $q$}$, where $\mbox{\boldmath $q$}$ is 
along the flow  direction. In Fig.\ref{fig.1¥}, 
 particles make longitudinal displacements (b) from the initial 
 uniform distribution (a). Similarly, particles make transverse displacements (c), 
 which is the origin of shear viscosity by transferring  
 the momentum from one streamline to another.  In reality, the 
 relative displacement in the flow is a mixture of these two types of 
 displacements $\delta\mbox{\boldmath $R$}=\delta\mbox{\boldmath 
 $R$}^l+\delta\mbox{\boldmath $R$}^t$. 

An important difference between these two types of displacements is as follows. 

(1)  Each vector $\delta\mbox{\boldmath $R$}^l$ of the longitudinal displacement
is parallel to the flow, and  therefore 
$\delta\mbox{\boldmath $R$}^l(\mbox{\boldmath  $x$}_i)$ of i-th 
particle in  $(\mbox{\boldmath $x$}_1,\cdots , \mbox{\boldmath $x$}_s)$ is added to
$\delta\mbox{\boldmath $R$}^l(\mbox{\boldmath $x$}_{i+1})$ of 
the neighboring particles on the streamline  as in Fig.\ref{fig.1¥}(b). Many 
$\delta\mbox{\boldmath $R$}^l(\mbox{\boldmath  $x$}_i)$
are accumulated in the final distribution $\delta\mbox{\boldmath 
 $R$}^l_{n+1}=\sum_{i=1}^{n¥}¥\delta\mbox{\boldmath $R$}^l_i$. 
 As a result, the longitudinal displacements induce an extensive 
 change in the particle distribution,
 and therefore, after the longitudinal displacement,  nearby particles do 
 not always exist in the initial distribution. Hence,  
$(\mbox{\boldmath $x$}_1+\delta\mbox{\boldmath $R$}^l(\mbox{\boldmath $x$}_1) ,\cdots , 
\mbox{\boldmath $x$}_s+\delta\mbox{\boldmath $R$}^l(\mbox{\boldmath  
$x$}_s))\equiv (\mbox{\boldmath $x$}_1^l,\cdots ,\mbox{\boldmath $x$}_s^l)$
cannot be reproduced by permutation of the initial 
$(\mbox{\boldmath $x$}_1,\cdots , \mbox{\boldmath $x$}_s)$, which 
means that the distance  between 
$(\mbox{\boldmath $x$}_1^l,\cdots ,\mbox{\boldmath $x$}_s^l)$ and
$(\mbox{\boldmath $x$}_1,\cdots ,\mbox{\boldmath $x$}_s)$ can be 
large in the configuration  space.

(2)  In contrast, $\delta\mbox{\boldmath $R$}^t$ of the 
transverse displacement is perpendicular to the flow, 
and therefore $\delta\mbox{\boldmath $R$}^t(\mbox{\boldmath  $x$}_i)$ 
of each particle is not additive as in Fig.\ref{fig.1¥}(c). 
Such transverse $\delta\mbox{\boldmath $R$}^t$ are not accumulated, 
thus no extensive change in the particle distribution is induced. 
For any given particle after displacement,    
a nearby particle always exists in the initial distribution.
Only by permutation of such particles, can the original distribution be 
reproduced.  In Bose statistics, owing to  permutation symmetry, one cannot 
distinguish the following two types of  particles after displacement: one 
moved from a near position by a short displacement vector, and  the other 
moved from a distant position by a long displacement vector. 
As long as such transverse displacements occur within the coherent wave 
function, we cannot distinguish the long and short displacements. 
 The distance between the initial and final distributions in 
configuration space is, after multiple permutations, finally reduced 
to a small distance having the order of the interatomic distance 
$r_0$ in the liquid.
For a given point $(\mbox{\boldmath $x$}_1,\cdots ,\mbox{\boldmath $x$}_s)$ 
in configuration space, the region that transverse 
$(\mbox{\boldmath $x$}_1+\delta\mbox{\boldmath $R$}^t(\mbox{\boldmath $x$}_1) ,\cdots , 
\mbox{\boldmath $x$}_s+\delta\mbox{\boldmath $R$}^t(\mbox{\boldmath  
$x$}_s))$ occupies is located much closer to this point than in the case of  
 longitudinal displacements. As $\mbox{\boldmath $q$}$ in  $\chi_{\mu\nu}(\mbox{\boldmath $q$}, \omega  )$ 
approaches zero in a boson's flow, this difference between the longitudinal and 
transverse displacements becomes more evident.

(b)  Let us consider the kinetic energy of such a system. 
Since the translational motion is possible only in the $z$-direction 
in a capillary, we define a collective coordinate of the translation as
\begin{equation}
 \xi _z=\frac{1}{s¥}¥\sum_{n=1}^{s¥}z_n .
	   	\label{eq:2953¥}
\end{equation}¥
The corresponding momentum $\pi _z$ is determined so that it 
satisfies a commutation relation $[\pi _z, \xi _z]=-i\hbar$ as
\begin{equation}
 \pi _z=-i\hbar¥\sum_{n=1}^{s¥}\frac{\partial}{\partial z_n¥} .
	   	\label{eq:2955¥}
\end{equation}¥
From the kinetic energy of the many-body system
\begin{equation}
 T=\frac{\hbar ^2}{2m¥}¥\sum_{n=1}^{s¥}¥\left(¥\frac{\partial ^2}{\partial x_n^2¥}¥
   +\frac{\partial ^2}{\partial y_n^2¥}¥+\frac{\partial ^2}{\partial z_n^2¥}¥\right)
 \psi(\mbox{\boldmath $x$}_1,\cdots ,\mbox{\boldmath $x$}_s)¥ ,
	   	\label{eq:302¥}
\end{equation}¥
we can extract the kinetic energy of the translational motion as a 
collective mode \cite {tom}, with the result that
 \begin{eqnarray}
 T&=&\frac{\hbar ^2}{2M¥}¥\left(¥\sum_{n=1}^{s¥}¥\frac{\partial }{\partial z_n¥}¥
 \psi(\mbox{\boldmath $x$}_1,\cdots ,\mbox{\boldmath $x$}_s)¥\right)^2 
                   \nonumber \\
 &+&\frac{\hbar ^2}{2m_{ef}¥}¥\sum_{n=1}^{s¥}¥\frac{\partial ^2}{\partial z_n^2¥}¥
 \psi(\mbox{\boldmath $x$}_1,\cdots ,\mbox{\boldmath $x$}_s)¥ 
                 \nonumber \\
 &+& \frac{\hbar ^2}{2m¥}¥\sum_{n=1}^{s¥}¥\left(¥\frac{\partial ^2}{\partial x_n^2¥}¥
			  + \frac{\partial ^2}{\partial y_n^2¥}¥\right)
 \psi(\mbox{\boldmath $x$}_1,\cdots ,\mbox{\boldmath $x$}_s)¥ .
	   	\label{eq:304¥}    
\end{eqnarray}¥
The first term is the kinetic energy of the translational motion 
responsible for $A(T)\delta (\omega)$ in the fluidity [Eq.(\ref{eq:2983¥})], 
and the second and third terms are those of the internal motion responsible for $\sigma_s(\omega)$.
($M=sm$ is the total mass of $s$ particles, and $m_{ef}$ is the effective mass 
of the internal motion along the $z$-direction.)

In general, the excited-state  wave function
$\psi(\mbox{\boldmath $x$}_1,\cdots ,\mbox{\boldmath $x$}_s)$ is orthogonal to the 
ground-state one  $\psi _0(\mbox{\boldmath $x$}_1,\cdots ,\mbox{\boldmath $x$}_s)$.
Since the latter has a constant amplitude,
 the former must spatially oscillate between the positive and negative values. 
Since the distance  between the longitudinal displacement 
$(\mbox{\boldmath $x$}_1^l,\cdots ,\mbox{\boldmath $x$}_s^l)$ and the 
initial one $(\mbox{\boldmath $x$}_1,\cdots ,\mbox{\boldmath $x$}_s)$ along the 
$z$-direction can be large, the derivatives of the wave function 
$\partial\psi/\partial z_i$ or $\partial ^2\psi/\partial z_i^2$ do 
not have a large value.  The kinetic energy of the longitudinal motion
[first and second terms in Eq.(\ref{eq:304¥})]
can have an infinitely small value, and a gap does not appear in its excitation spectrum.

 On the other hand, since a large transverse displacement is not 
 possible owing to permutation symmetry, the distance between 
 the transverse  displacement 
$(\mbox{\boldmath $x$}_1^t,\cdots ,\mbox{\boldmath $x$}_s^t)$ and the 
initial one $(\mbox{\boldmath $x$}_1,\cdots ,\mbox{\boldmath $x$}_s)$ 
is not large. Hence, the derivatives of the wave function 
$\partial ^2\psi/\partial x_i^2$ or $\partial ^2\psi/\partial y_i^2$ do 
not have a small value.  Owing to a steep increase and decrease in amplitude within a short 
distance, the kinetic energy of the internal motion [third term in 
Eq.(\ref{eq:304¥})] increases the transverse excitation energy  \cite {fey1}.  
As a result, the transverse motion within the coherent wave function 
is suppressed by this statistical gap having the order of 
$\epsilon_0=(\hbar^2/2m_t)(1/r_0^2)=\hbar a_t/r_0^2$  
 as in Eq.(\ref{eq:053¥}), which explains the rigidity of the coherent 
 wave function.  To put it simply, the kinetic energy of the 
 superfluid flow is composed only of the first and second terms in 
 Eq.(\ref{eq:304¥})  \cite {tak}.

In summary, during the growth of a superfluid flow,  permutation 
symmetry by Bose statistics plays a role similar to the long-range interaction,
while the repulsive short-range interaction $U$ determines the 
interatomic distance $r_0$.  Both interactions 
cooperate to realize the macroscopic superfluid flow.

(c) The statistical gap changes the shear viscosity as follows. In the 
fluctuation-dissipation theorem, the shear viscosity coefficient $\eta$ \cite {mor}
of a flowing liquid along the $z$-axis is represented  by
\begin{equation}
 \eta = \frac{1}{k_BTV¥}¥\int_{0}^{\infty¥}¥\langle  S^{rz}(t)S^{rz}(0)\rangle  dt,
	   	\label{eq:305¥}
\end{equation}¥
where $S^{rz}$ stands for the stress tensor
\begin{equation}
   S^{rz}(t) = \sum_{i}^{N¥}¥\frac{1}{m¥}¥(p^x_i+p^y_i)p^z_i + \frac{1}{2¥}¥ 
   \sum_{i}^{N¥}¥\sum_{j}^{N¥}r_{ij}\frac{\partial U}{\partial 
   z_{ij}¥}¥,
	   	\label{eq:306¥}
\end{equation}¥
where the second term is the effect of the repulsive interaction.
The transverse motion along the direction perpendicular to the flow in the condensate is 
suppressed by the statistical gap, and  the resulting zero  
transverse-momenta $p^x_i$ and $p^y_i$ of each particle cause the 
shear viscosity $\eta$ to vanish.  The fluidity of a superfluid flow with such $\eta =0$ 
is given by  $A(T)\delta (\omega)$ in Eq.(\ref{eq:02¥}). 

  \begin{figure}
		\begin{center}
\includegraphics [scale=0.5]{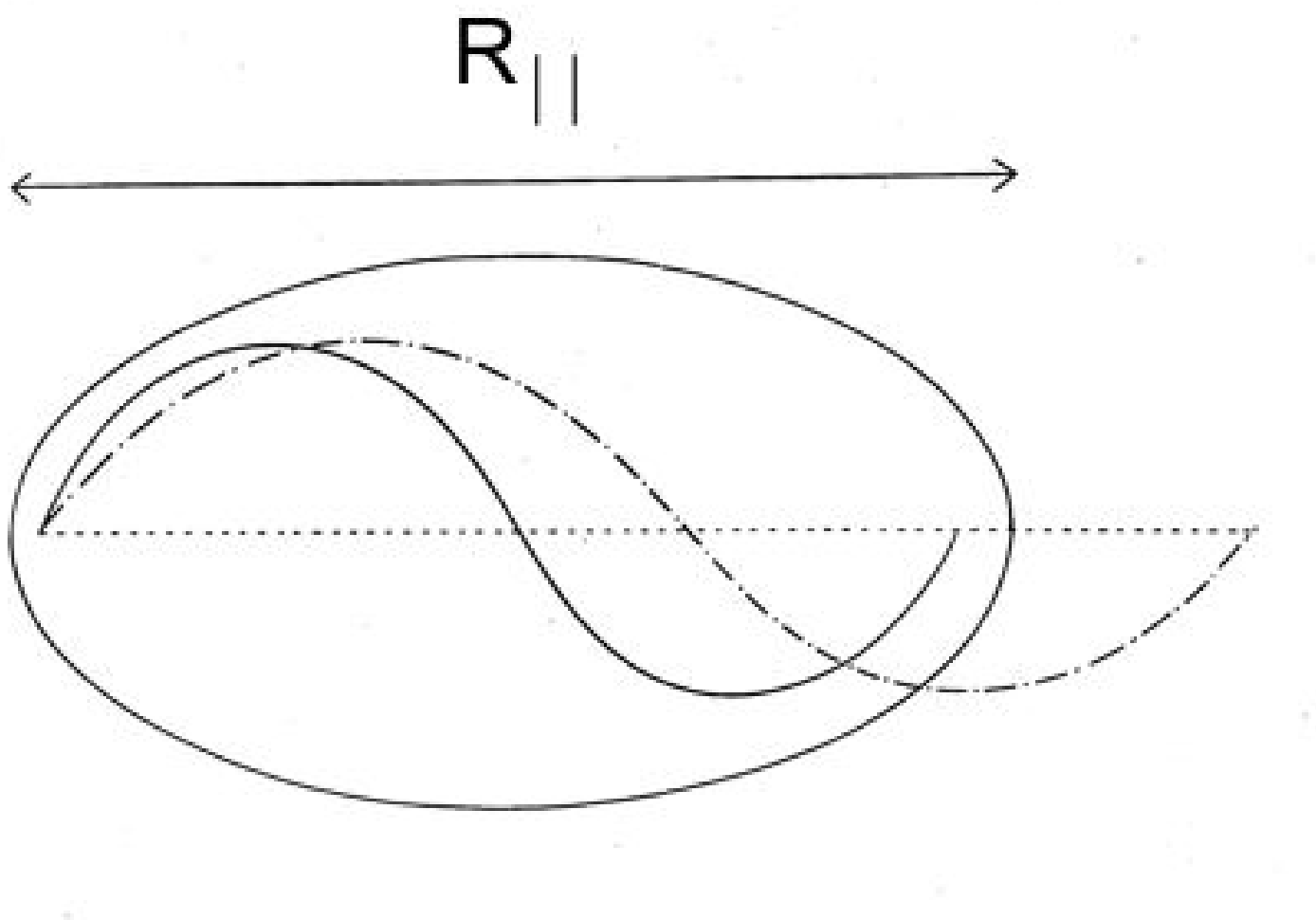}
\caption{\label{fig:epsart}  Average form of the coherent many-body 
wave function shown as an ellipsoid with major axis $R_{\parallel }$ 
along the flow. The solid curve represents the transverse displacement 
from the horizontal dotted  straight line within the 
coherent wave function, and the one-point-dotted curve represents that beyond the coherent 
wave function. For a given $\omega$, the critical size $s_1$ of the 
coherent wave function for distinguishing the above two cases is given by Eq.(\ref{eq:353¥}). }
\label{fig.055¥}
     	\end{center}
\end{figure}

(d) For $\sigma _s(\omega)$ in Eq.(\ref{eq:02¥}), the statistical gap
in the transverse motion strongly affects the spectrum of fluidity
as follows. The angular frequency $\omega$  in $\sigma_1(\omega)$ corresponds  to
the time scale at which we observe the dynamics of a flow. 
 In this case, the transverse excitation energy $\epsilon _t(s,q)$ depends on whether 
 the wavelength $\lambda=2\pi/q$ of the excitation exceeds the length $R_{\parallel }$  of the coherent 
wave function, as illustrated in Fig.\ref{fig.055¥}. 
If the one-particle transverse excitation occurs  
within the coherent wave function, it is suppressed by the 
statistical gap as $\epsilon _t(s,q)=\epsilon _0$, but if the excitation occurs
beyond this wave function, permutation symmetry does not hold for particles involved
in this excitation, and it has an ordinary one-particle 
transverse-excitation energy $\epsilon _t(s,q)=(\hbar q)^2/2m_t=\hbar a_tq^2$. 
Hence, to study the growth process of a superfluid flow, we must  take into account
the size distribution of the coherent many-body wave function in the flow.

\subsection {Size distribution of coherent many-body wave function}
  In a condensate near $T_{\lambda}¥$, there are large or small coherent wave functions.
To estimate the effect of the statistical gap  $\epsilon _0$ on the 
flow, we need the size distribution $h(s)$ of the coherent wave function
in the flow. Here, we simply approximate the Hamiltonian as $H=\sum [\epsilon_p+U_0]¥$, 
where the constant $U_0$ is the mean-field value of the 
repulsive interaction energy per particle. The grand partition 
function of bosons in local equilibrium $Z_0(\mu)=\prod(1-e^{-\beta(\epsilon_p 
+U_0-\mu_0)}¥)^{-1}$, where $\mu_0(T)$ is the chemical potential of the 
ideal  boson gas, is rewritten  as 
\begin{equation}
	 Z_0(\mu) =\exp \left[-\sum_p \ln(1-ze^{-\beta\epsilon _p })¥\right]¥,
	   \label{eq:05¥}
\end{equation}¥
where  $z=\exp (\beta[\mu_0-U_0])$.   We define a 
new $\mu (T)$  as  $\mu _0(T)$ including $U_0$ 
\begin{equation}
   	 \mu (T) = -\left(\frac{g_{3/2}(1)\times 3}{4\sqrt{\pi}¥}¥\right)^2
	 k_BT_{\lambda}\left(\frac{T-T_{\lambda}}{T_{\lambda}¥}\right)^{2}-U_0 ,
  	        \label{eq:055¥}
\end{equation}¥ 
where $g_{3/2}(1)=\sum_{n=1}^{\infty¥}¥n^{-3/2}=2.612$ (Appendix.B).
With decreasing temperature, the  chemical  potential $\mu_0(T)$ of 
the ideal Bose gas  approaches zero,
but  $U_0$ is a finite quantity in the liquid.  
By changing $U_0$, we can control how many particles participate in the 
condensate at $T_{\lambda}¥$.

In the grand canonical ensemble, the particle number $s$ appears in 
the energy as $\mu s$. Hence, if we expand $Z_0(\mu)$ in 
powers of $z=\exp (\beta\mu)$, it  yields the size distribution of the coherent wave 
function. The exponent in Eq.(\ref{eq:05¥}) for $p=0$ and $p \neq 0$ is expanded in powers 
of $z$ as
\begin{equation}
    \ln(1-z) = -\sum_{s=1}^{\infty}\frac{z^s}{s¥} ,	
	\label{eq:06¥}
\end{equation}¥
for $p =0$, and expanded as
\begin{equation}
     \frac{4}{\sqrt{\pi }}\int_{0}^{\infty¥}x^2\ln(1-ze^{-x^2})dx¥
    = -\sum_{s=1}^{\infty}\frac{z^s}{s^{5/2}¥},	
	\label{eq:07¥}
\end{equation}
for $p \neq 0$, where $\sum_{p}¥$ in Eq.(\ref{eq:05¥}) is transformed to an integral over
$x=\sqrt{\beta\epsilon _p}¥$.  With these expansions
used in Eq.(\ref{eq:05¥}), we obtain the following $Z_0(\mu)$ \cite {fey2}:
\begin{equation}
	 Z_0(\mu) =\exp \left[V\sum_{s=1}^{\infty ¥}
	             \left(\frac{e^{\beta\mu s}}{s¥}
	    +\frac{V}{2\lambda _t^3}\frac{e^{\beta \mu s}}{s^{5/2}¥}\right)¥\right]¥,
	   \label{eq:08¥}
\end{equation}¥
where $\lambda _t=\sqrt{2\pi\hbar^2\beta/m}$ is the thermal wavelength. 
 The equation of states $N=k_BT\partial \ln Z_0/\partial\mu$ 
  with this $Z_0(\mu) $ is written  as a sum over the size $s$ of the 
 coherent many-body wave function  
\begin{equation}
 \frac{N}{V¥}¥=\sum_{s=1}^{\infty ¥}
	   \left(e^{\beta\mu s}
	     +\frac{V}{2\lambda _t^3}\frac{e^{\beta\mu s}}{s^{1.5}¥}\right)¥ .
      \label{eq:09¥}
\end{equation}

\begin{figure}
		\begin{center}
\includegraphics [scale=0.6]{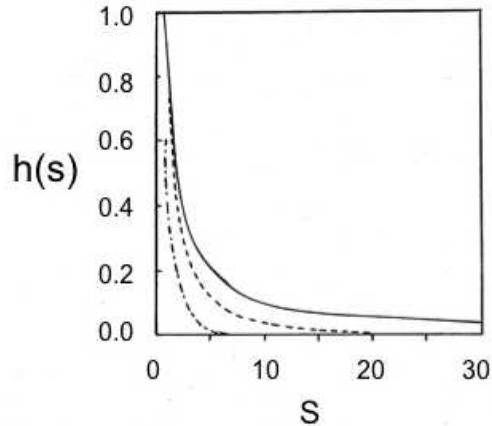}
\caption{\label{fig:epsart} Size distribution $h(s)=\exp (\beta\mu s)/s$ 
for $p=0$ of the coherent many-body wave functions at different 
temperatures: $h(s)$ at $1.5T_{\lambda}$ (one-point dotted line), at $1.2T_{\lambda}$
(dotted line), and at $T_{\lambda}$  (solid line). }  
\label{fig.2¥}
     	\end{center}
\end{figure}

In Eq.(\ref{eq:09¥}), each term on the right-hand side represents the 
number density of particles belonging to the coherent many-body wave function 
including $s$ particles. (For the combinatorial derivation of 
Eq.(\ref{eq:08¥}), see Appendix C.)
Dividing each term in Eq.(\ref{eq:09¥}) by $s$, 
one obtains the size distribution $h(s)$ of the coherent many-body wave function  

\begin{equation}
 h(s) = \displaystyle{\frac{\exp (\beta\mu s)}{s¥}¥},
	       \qquad   p=0
	     	\label{eq:24102¥}
\end{equation}¥
\begin{equation}
 h(s) =\displaystyle{ \frac{V}{2\lambda_t^3¥}¥
		\frac{\displaystyle{\exp (\beta\mu s)}}{s^{2.5}¥}¥},      ¥
	       \qquad   p\ne 0   
	     	\label{eq:24104¥}
\end{equation}¥


 in analogy with the cluster expansion  of the equation of states in a 
classical gas.  Figure \ref{fig.2¥} shows  $h(s)$ for $p=0$. As the
temperature approaches  $T_{\lambda}$, it  changes from a rapidly 
falling function of $s$ at $1.5T_{\lambda}$ to a function extending to  
large $s$ at $T_{\lambda}$.  As will be explained in Sec.4, 
the lower-convex form of $h(s)$ in Fig.\ref{fig.2¥} translates to
the lower-convex form of $\sigma_1(\omega)$ of superfluid flow in 
Fig.\ref{fig.015¥} through  the Kramers--Kronig relation. 
While the left-hand side of Eqs.(\ref{eq:06¥}) and  (\ref{eq:07¥}) is well known
as the one-particle distribution of Bose statistics, the right-hand side is useful
for the argument regarding the size of the coherent wave function.

Apart from this $h(s)$, we do not have sufficient information 
on the shape of the coherent wave function in the flowing liquid.  However, it is 
natural to assume that its average shape is a compact form, which is
symmetric with respect to the rotation around  the direction of flow. 
A compact form with such a symmetry is the ellipsoid 
$(x^2+y^2)/R_{\perp }^2+z^2/R_{\parallel }^2=1 $, 
as illustrated in Fig.\ref{fig.055¥}, with 
 major axis $R_{\parallel }$ (parallel to the flow along the $z$-axis) 
 and minor one $R_{\perp }$ (on the $x-y$ plane perpendicular to the flow). 
 The number $s$ of particles included in the coherent wave 
function is proportional to the  
 volume of the ellipsoid as $(4\pi/3)R_{\parallel }R_{\perp }^2=sr_0^3$. 
 The shape of the ellipsoid (the ratio of $R_{\parallel }$ to 
 $R_{\perp }$) depends on $s$,   which is 
 represented by $R_{\parallel }\simeq s^Zr_0$ and $R_{\perp }\simeq s^Xr_0$ with $Z+2X=1$.  
When the condensate grows in a liquid at rest, its average shape is a sphere with $Z=1/3$. 
However, when the coherent wave function appears  in a flowing 
 liquid at a low temperature, there is already a velocity distribution 
$\mbox{\boldmath $v$}(\mbox{\boldmath $r$})$, as shown in  Fig.\ref{fig.1¥}, and therefore,
the coherent wave function, which rides on this $\mbox{\boldmath $v$}(\mbox{\boldmath $r$})$,
is elongated  along the flow direction ($z$-axis). The ellipsoid is likely to be more
 elongated  as $s$ increases.  In reality, the average shape of the coherent 
many-body wave function in the flow is likely to have a value of $Z$ near to 1.

In Sec.3, we prepare the linear-response formalism for the fluidity 
spectrum $\sigma_1(\omega)$, and in Sec.4, we will combine $h(s)$ in Fig.\ref{fig.2¥}
with the image  in  Fig.\ref{fig.055¥}  using the formalism prepared in 
Sec.3.

\section {Fluidity Spectrum and Translational Superfluid Flow.  }

\subsection  {Fluidity spectrum as a linear-response coefficient}
Let us consider a repulsive Bose system with the following Hamiltonian
\begin{equation}
 H= \sum_{p} \epsilon (p)\psi^{\dagger}_{p}\psi _{p}
  + U\sum_{p,p'}\sum_{q}\psi^{\dagger}_{p-q}\psi^{\dagger}_{p'+q}\psi 
  _{p'}\psi _{p} \qquad (U>0). 
  \label{eq:269¥}
\end{equation}¥
 The fluidity spectrum $\sigma (\omega)$ in Eq.(\ref{eq:01¥}) can 
be generalized from a real to a complex number 
\begin{equation}
 \mbox{\boldmath $j$}(\omega)=\left[\sigma_1(\omega)+i\sigma_2(\omega)\right]¥d^2 
 \frac{\mbox{\boldmath $P$}(\omega)}{L¥}.
	     	\label{eq:2411¥}
\end{equation}¥
In contrast to the mechanical perturbation which is directly 
exerted on particles such as the electric field on electrons, the pressure difference $P$ 
between the two ends of a capillary is a thermal perturbation, 
which is a far more complex phenomenon.  Among some formalisms 
dealing with this problem \cite{zwa}, we will adopt the simplest one called
the {\it indirect method \/}. 
To incorporate Eq.(\ref{eq:2411¥}) into the linear-response theory, 
 we will define, instead of $P(t)=P(\omega)\exp(-i\omega t)$, 
 a fictitious momentum-like field  $Q(t)=Q(\omega)\exp(-i\omega t)$ 
 satisfying the following equation of motion  \cite{sup}  
\begin{equation}
   \frac{dQ(t)}{dt¥}¥=  \frac{P(t)}{L¥}¥ .
	     	\label{eq:2415¥}
\end{equation}¥
This $Q(\omega)$ can be regarded as constituting the 
perturbation energy  as 
\begin{equation}
      \frac{1}{\rho¥}¥\int ¥j(\omega)Q(\omega)d\omega .
	     	\label{eq:242¥}
\end{equation}¥
Instead of $P(\omega)$, we can formally regard $Q(\omega)$ as an external field 
inducing $j(\omega)$ in Eq.(\ref{eq:2411¥}).  Here, we make an 
approximation that this $\mbox{\boldmath $Q$}(\omega)$ is instantaneously 
exerted on the flow $\mbox{\boldmath $j$}(\omega)$ similarly to the true 
mechanical one.  Since a liquid is a system with high density, the 
effect of pressure, represented by $\mbox{\boldmath $Q$}(\omega)$,
 rapidly propagates through the liquid in a capillary.
 Compared with a  gas in which such a propagation requires 
considerable time, this approximation is not an unrealistic one 
 in a liquid, especially for the slow perturbation.  

We obtain a bosonic analogue of the London equation in superconductivity by rewriting 
Eq.(\ref{eq:2411¥}), using $P(\omega)/L=-i\omega Q(\omega)$ of 
Eq.(\ref{eq:2415¥}), as 
\begin{equation}
 \mbox{\boldmath $j$}(\omega)
         =\left[\omega\sigma_2(\omega)-i\omega\sigma_1(\omega)\right]¥¥d^2
                                        \mbox{\boldmath $Q$}(\omega),
             	\label{eq:25¥}
\end{equation}¥
and regard $\left[\omega\sigma_2(\omega)-i\omega\sigma_1(\omega)\right]¥¥d^2$ 
as a spatially  averaged form of the generalized susceptibility. 
In this formula, one can regard the real part $\omega\sigma_2(\omega)$  
as representing the nondissipative process, and the imaginary part 
 $-\omega\sigma_1(\omega)$ as representing the dissipative one.

The fluctuation-dissipation theorem says that the nonequilibrium 
behavior of the system, which is slightly deviated from the 
equilibrium state, is described by the generalized susceptibility 
$\chi_{\mu\nu}(\mbox{\boldmath $q$},\omega )$ of the equilibrium state. 
Microscopically, the capillary flow of a superfluid is described by the 
generalized susceptibility composed of the flow-flow correlation.
\begin{equation}
	\chi_{\mu\nu}(\mbox{\boldmath $q$}, i\omega _n )
 = -\frac{1}{V¥}¥\int_{0}^{\beta¥}¥d\tau\exp ( i\omega _n \tau)
  \langle 0|T_{\tau}J_{\mu}(\mbox{\boldmath $q$},\tau)J_{\nu}(\mbox{\boldmath $q$},0)|0\rangle , 
  \label{eq:27¥}
\end{equation}¥
where
\begin{equation}
J_{\mu}(\mbox{\boldmath $q$},\tau)= \sum_{n}\sum_{p}  \hbar 
\left(p+\frac{q}{2¥}¥\right)_{\mu} \psi^{\dagger}_{p}\psi _{p+q}\exp ( i\omega _n \tau). 
  \label{eq:275¥}
\end{equation}¥
Behind the macroscopic velocity $v(r)$ in Eq.(\ref{eq:005¥}), the microscopic velocity 
of each particle has not only longitudinal but also transverse components. Accordingly,
the generalized susceptibility consists of the longitudinal and transverse parts ($\mu, \nu =x,y,z$) 
\begin{equation}
	\chi_{\mu\nu}(\mbox{\boldmath $q$},\omega )
	      =\chi^L(\mbox{\boldmath $q$},\omega)\frac{q_{\mu}q_{\nu}}{q^2¥}     
	       +\chi^T(\mbox{\boldmath 
		   $q$},\omega)\left(\delta_{\mu\nu}-\frac{q_{\mu}q_{\nu}}{q^2¥}\right)¥ .
	   \label{eq:265¥}
\end{equation}¥
 In a rotational superfluid flow, $\chi_{\mu\nu}(\mbox{\boldmath $q$},\omega )$ 
 describes the response to the rotation of the container, and 
 $\lim _{q\rightarrow 0}[\chi^L(\mbox{\boldmath $q$},0)-\chi^T(\mbox{\boldmath $q$},0)¥]$ 
  is regarded as the superfluid density $\rho _s$.  (Historically, it is traced back to Schafroth 
in the study of superconductivity before the advent of BCS theory \cite {sch}.)
When we consider the translational superfluid flow, in view of the 
continuity to the rotational superfluidity, we will begin with the 
form $\sigma(\omega =0)=\rho/(4\eta)
=\lim _{q\rightarrow 0}[\chi^L(\mbox{\boldmath $q$},0)-\chi^T(\mbox{\boldmath $q$},0)¥]/(4\eta)$. 

The Kramers--Kronig relation connects the real and imaginary parts of 
susceptibility.  In the transverse response by $\chi^T(\mbox{\boldmath $q$},\omega)$, 
the particle displacements are orthogonal to the external force, 
thus resulting in the nondissipative response $\omega\sigma_{2}(\omega)$ in Eq.(\ref{eq:25¥}).
Hence, the nondissipative part  $\omega\sigma_2(\omega)=\omega\rho/(4\eta)$,
where $\omega =aq^2$, is represented by 
$a\int dq^2¥\chi^T(\mbox{\boldmath $q$},\omega)/(4\eta _n)$. 
(Since we are interested in not only the macroscopic 
but also a mesoscopic condensate, we integrate 
$\chi^T(\mbox{\boldmath $q$},\omega)$ over $q$, instead of taking $q \rightarrow 0$.)  
On the other hand, the longitudinal  response by $\chi^L(\mbox{\boldmath $q$},\omega)$
corresponds to the dissipative part  $\omega\sigma_1(\omega)$.
The Kramers--Kronig relation for Eq.(\ref{eq:25¥})  is written as 
\begin{equation}
   -\omega\sigma_{1}(\omega)= -\frac{1}{\pi}¥¥\frac{1}{4\eta_n¥}¥
         \int_{-\infty}^{\infty¥}d\xi \int adq^2¥
		 \frac{\chi^T(\mbox{\boldmath $q$},\xi)}{\xi-\omega¥}¥.
	    	\label{eq:2673¥}
\end{equation}¥
 In the coarse-graining ($\mbox{\boldmath $q$}\rightarrow 0$) of a normal liquid, 
 the response to the external force with a larger 
wavelength becomes more  irrelevant to its microscopic  structure.
Hence, it does not matter in which direction the external force is 
exerted on the liquid  at $\mbox{\boldmath $q$}\rightarrow 0$, 
with the result that $\lim _{q\rightarrow 0}\chi^L(\mbox{\boldmath $q$},\omega)
=\lim _{q\rightarrow 0}\chi^T(\mbox{\boldmath $q$},\omega)$ at small 
$\omega$.  Instead of Eq.(\ref{eq:2673¥}), in analogy to the motion of a rigid body, 
we usually define the fluidity  $\sigma_{1n}(\omega)$ of a normal fluid as 
\begin{equation}
   \sigma_{1n}(\omega)= \frac{1}{\pi\omega}¥¥\frac{1}{4\eta_n¥}¥
         \int_{-\infty}^{\infty¥}d\xi \int adq^2¥
		 \frac{\chi^L(\mbox{\boldmath $q$},\xi)}{\xi-\omega¥}¥.
	    	\label{eq:281¥}
\end{equation}¥

\subsection  {Two-fluid model in terms of fluidity spectrum }

In a superfluid, the coherent wave function grows to 
a macroscopic size, and therefore a difference between the 
longitudinal  and transverse properties continues to exist even 
at $\mbox{\boldmath $q$}\rightarrow 0$, 
with the result that $\lim _{q\rightarrow 0}\chi^L(\mbox{\boldmath $q$},\omega)
\neq\lim _{q\rightarrow 0}\chi^T(\mbox{\boldmath $q$},\omega)$.
Accordingly, instead of Eq.(\ref{eq:281¥}), one must go back  to 
the original definition  [Eq.(\ref{eq:2673¥})] so that it is 
applicable to a superfluid.
For a superfluid, we  rewrite Eq.(\ref{eq:2673¥}) as \cite{noz}
 \begin{eqnarray}
\sigma_1(\omega)&=&\sigma_{1n}(\omega) 
			        \nonumber   \\     
		 &-&\frac{1}{\pi \omega¥}¥\frac{1}{4\eta_n¥}¥\int_{-\infty}^{\infty¥}d\xi 
           \int adq^2¥ \frac{\chi^L(\mbox{\boldmath	$q$},\xi)
			    -\chi^T(\mbox{\boldmath 
				$q$},\xi)¥}{\xi-\omega¥}¥
				\nonumber   \\   
		&+&A(T)\delta(\omega). 
				 \nonumber   \\ 
	   	\label{eq:29¥}
\end{eqnarray}¥
The second term represents the transition from 
 the normal flow to the superfluid flow.  We will obtain 
$\chi^L(\mbox{\boldmath	$q$},\xi)-\chi^T(\mbox{\boldmath $q$},\xi)¥$ 
using the quantum mechanical many-body theory. We integrate 
$\chi^L(\mbox{\boldmath $q$},\omega) -\chi^T(\mbox{\boldmath $q$},\omega)¥$
  over all possible $\mbox{\boldmath $q$}$  at a given $\omega$ in Eq.(\ref{eq:29¥}). 
 The third term represents the fluidity of a macroscopic superfluid flow. 
 In a translational superfluid 
 flow, the fluidity is a quantity determined by the fundamental parameters  rather than the 
 phenomenological ones such as $\eta_n$. In the simplest two-fluid model, one considers
 only $\sigma_{1n}(\omega) $ and $A(T)\delta(\omega)$ in $\sigma_{1}(\omega) $.

 This is the revised expression of the two-fluid model in terms of 
 the fluidity spectrum.  By the spectral analysis of how such a 
 $\chi^L(\mbox{\boldmath $q$},\omega) -\chi^T(\mbox{\boldmath $q$},\omega)$ 
gradually grows with decreasing temperature, we can explore the formation of 
the translational superfluid flow, which is not seen in the rotational 
superfluid flow  \cite {busk} (Appendix D).
 At the  $q\rightarrow 0$ limit, the longitudinal excitation from the 
condensate is equivalent to the uniform translation of the condensate itself.

\section { Physical Model of Translational Superfluid Flow }
On the basis of the physical view in Sec 2, let us consider  the fluidity spectrum 
$\sigma_1(\omega)$ of the superfluid flow in liquid. 

 (1)   In a slowly moving Bose system in which the physical quantities 
slowly  vary in space and time,  we can regard the system as a mixture of some local 
equilibrium states, and therefore  the first approximation of
$\chi_{\mu\nu}(\mbox{\boldmath $q$}, \omega )$ 
 is one-particle excitations from the condensate (the  zeroth order of $U$) given by 
  \begin{eqnarray}
   	\chi_{\mu\nu}(\mbox{\boldmath $q$}, \omega  )
	=&-& \sum_{p} \hbar^2\left(p+\frac{q}{2¥}¥\right)_{\mu}\left(p+\frac{q}{2¥}¥\right)_{\nu}
		       \nonumber\\ 
 &\times &	\frac{f_B(p)-f_B(p+q)}{\hbar\omega +\epsilon (p)-\epsilon (p+q)¥}¥ , 
  \label{eq:277¥}
\end{eqnarray}¥
where $f_B(p)$ is the Bose distribution function.

In view of $\chi_{\mu\nu}(\mbox{\boldmath $q$},\omega )$ in Eq.(\ref{eq:265¥}), 
$\chi^L(\mbox{\boldmath $q$},\omega)-\chi^T(\mbox{\boldmath 
$q$},\omega)¥$ is the coefficient of $q_{\mu}q_{\nu}/q^2¥$. Hence, in 
the case of  Eq.(\ref{eq:277¥}), we obtain
 \begin{eqnarray}
   [\chi^L(\mbox{\boldmath $q$},\omega)&-&\chi^T(\mbox{\boldmath 
	$q$},\omega)]   \frac{q_{\mu}q_{\nu}}{q^2¥}¥ 
	           \nonumber\\ 
 &=&-\frac{ \hbar^2}{4¥}¥¥\sum_{p}
	\frac{f_B(p)-f_B(p+q)}{\hbar\omega +\epsilon (p)-\epsilon (p+q)¥}¥q^2
		 \times\frac{q_{\mu}q_{\nu}}{q^2¥}¥ . 
		       \nonumber\\ 
  \label{eq:282¥}
\end{eqnarray}¥

 When viewing the superfluid flow at $T=0K$ on a reference frame moving at a 
 momentum $p$, the system consists of various   
coherent wave functions having zero momentum ($p=0$), and
we can simplify the sum over  $p$ by putting $f_B(p)=0$ for a finite $p$.  
  While a capillary flow is originally a dissipative process, time-reversal symmetry  holds 
for $\chi^L(\mbox{\boldmath $q$},\omega)-\chi^T(\mbox{\boldmath 
$q$},\omega)¥$ in the Kramers--Kronig relation of Eq.(\ref {eq:29¥}). 
Hence, we consider in Eq.(\ref{eq:282¥}), not only a term with $p=0$ 
representing an excitation $\epsilon(0) \rightarrow \epsilon(q)$ from the condensate,  
but also a term with $p=-q$  representing a decay  $\epsilon(-q) 
\rightarrow \epsilon(0)$ to the condensate. 
The former corresponds to  $\epsilon (p)-\epsilon 
(p+q)\simeq -\hbar^2q^2/(2m) \equiv -\hbar aq^2$, where $a=\hbar/(2m)$, 
and the latter to  $\epsilon (p)-\epsilon 
(p+q)\simeq \hbar^2q^2/(2m)= \hbar aq^2$. Hence, we find that
\begin{equation}
  \chi^L(\mbox{\boldmath $q$},\omega)-\chi^T(\mbox{\boldmath $q$},\omega)¥
     =-\frac{\hbar}{4¥}¥ 
	 \left(\frac{f_B(0)}{\omega-aq^2¥}¥+\frac{-f_B(0)}{\omega+aq^2¥}¥\right)¥q^2 .
                   \label{eq:284¥}
\end{equation}¥
(When $\omega$ is defined in $-\infty<\omega<\infty$, the first and 
second terms have the same form.)

(2) To improve $\chi_{\mu\nu}(\mbox{\boldmath $q$},\omega )$ so that 
it reflects the situation of the liquid, 
we  construct the longitudinal and transverse susceptibilities 
 by comparing Eqs.(\ref{eq:265¥}) and (\ref{eq:277¥}). 
 Since the transverse excitation energy depends on 
 the size of the coherent wave function, we decompose $f_B(p=0)$ in Eq.(\ref{eq:284¥})
 into a sum over $s$ using  $h(s)=\exp (\beta\mu s)/s$ for  $p=0$ \cite {unit}.  
Such a $\chi_{\mu\nu}(\mbox{\boldmath $q$},\omega )$ at a low frequency 
in $-\infty<\omega<\infty$ is given by 
 \begin{eqnarray}
	\chi_{\mu\nu}(\mbox{\boldmath $q$},\omega )
 &=& \sum_{s}^{\infty¥}¥sh(s)¥ \frac{\hbar}{2¥}¥\frac{-q^2}{\omega-a_lq^2¥}¥\frac{q_{\mu}q_{\nu}}{q^2¥}¥
						\nonumber\\   	 
            &+&  \sum_{s}^{\infty¥}¥sh(s)¥ \frac{\hbar}{2¥}¥\frac{-q^2}{\omega-\epsilon_t(s,q)¥/\hbar}¥
		  \left(\delta_{\mu\nu}-\frac{q_{\mu}q_{\nu}}{q^2¥}\right)¥ ,
		                	\nonumber\\  
							\label{eq:2995¥}
\end{eqnarray}¥
where $q$ is mainly a wave vector along the flow direction.  
The excitation energy $\epsilon_t(s,q)$ of the transverse motion 
depends on the size $s$ of the coherent wave function: 
(a) $\epsilon_t(s,q)=\epsilon_0$ for excitations within the coherent wave 
function as in Eq.(\ref{eq:053¥}), and (b) $\epsilon_t(s,q)=\hbar a_tq^2$ 
for that beyond the coherent wave function as shown in Fig.\ref{fig.055¥} \cite {dam}. 
We  distinguish the longitudinal one-particle excitation energy
$\epsilon(p+q)-\epsilon(p)=\hbar^2/(2m_l)\times q^2=\hbar a_lq^2$ 
 from that of the  transverse one $\hbar^2/(2m_t)\times q^2= \hbar a_tq^2$.

 A new form of $\sigma_1(\omega)$ is obtained  using 
Eq.(\ref{eq:2995¥}) as follows. 
Extracting $\chi^L(\mbox{\boldmath $q$},\omega)-\chi^T(\mbox{\boldmath $q$},\omega)¥$ 
from Eq.(\ref{eq:2995¥}) by comparing it with Eq.(\ref{eq:265¥}), and 
substituting this result in Eq.(\ref{eq:29¥}), we obtain the fluidity spectrum
 \begin{eqnarray}
 \sigma_1(\omega) = \sigma_{1n}(\omega) 
         &+&\frac{\hbar}{2\pi¥}\frac{1}{4\eta_n¥}¥
		  a\int dq^2\frac{q^2}{\omega¥}¥
		                     \sum_{s}^{\infty¥}¥sh(s)¥ \int_{-\infty}^{\infty¥}d\xi 
						\nonumber\\   	 
            &\times&  \frac{1}{\xi-\omega¥}¥
			\left(\frac{1}{\xi-a_lq^2¥}-\frac{1}{\xi-\epsilon_t(s,q)¥/\hbar}¥\right)¥
			        \nonumber\\     
			&+& A(T)\delta(\omega).
	   	\label{eq:2997¥}
\end{eqnarray}¥
Defining a new variable $x=\xi-aq^2$ in the 
above integral, we can use the Hilbert transformation of $1/x$, which 
is  another form of $1/(x+i\epsilon)=1/x-i\pi\delta(x)$, 
\begin{equation}
  \frac{1}{\pi¥}¥\int_{-\infty¥}^{\infty¥}¥¥\frac{dx}{x-\omega¥}¥
                      \frac{1}{x¥} =\pi ¥\delta(\omega)  .
	   	\label{eq:31¥}
\end{equation}¥
Applying the Hilbert transformation, we obtain
 \begin{eqnarray}
 \sigma_1(\omega)&=&\sigma_{1n}(\omega) 
						\nonumber\\   	 
      &+&\frac{\pi\hbar}{2¥}\frac{1}{4\eta_n¥}¥ a\int dq^2 
		   	\nonumber\\   
   &\times& 	\sum_{s}^{\infty¥}¥sh(s)¥\frac{q^2}{\omega¥}¥ 
			 \left[\delta(\omega-a_lq^2)-\delta(\omega-\epsilon_t(s,q)/\hbar)\right]¥
			 	\nonumber\\   
	&+& A(T)\delta(\omega).
	       	\label{eq:2999¥}      
\end{eqnarray}¥

 $\delta(\omega-a_lq^2)$ and $\delta(\omega-\epsilon_t(s,q)/\hbar)$ 
represent the contributions from the longitudinal and transverse 
excitations, being the result of the first and second 
terms in Eq.(\ref{eq:304¥}), and the third term in it, respectively. 
[$\sigma_{1n}(\omega ) $  
includes more complex phenomena than that in Eq.(\ref{eq:2995¥}).
However, we focus on the change in $\sigma_{1}(\omega ) $ near $T_{\lambda}$,
hence  assuming that this change is well described by the simple 
$\chi_{\mu\nu}(\mbox{\boldmath $q$},\omega )$  such as Eq.(\ref{eq:2995¥}).]

\subsection {$\sigma_1(\omega)$ from the transverse motion  }

According to the mechanism explained in Fig.\ref{fig.055¥},
the sum over $s$ in the transverse motion in Eq.(\ref{eq:2999¥}) is  
divided into two parts: a  
sum ($s>s_1$) in which excitations occur within the same coherent 
wave function and  the statistical gap  $\epsilon _0$ suppresses the transverse
 excitations, and a sum ($s<s_1$)  in which excitations occur  beyond 
 the coherent wave function and permutation symmetry no longer 
 restricts the transverse excitations of the system. 
  \begin{eqnarray}
\sum_{s=1}^{\infty¥}¥&s& h(s)\frac{q^2}{\omega ¥}¥\delta(\omega -\epsilon_t(s,q)/\hbar)
		   	\nonumber\\  
&=& \sum_{s=1}^{s_1¥}¥sh(s)\frac{q^2}{\omega ¥}¥\delta(\omega -a_tq^2)
    +\sum_{s=s_1+1}^{\infty¥}¥sh(s)\frac{q^2}{\omega ¥}¥\delta(\omega -\epsilon_0/\hbar) 
	          	\nonumber\\ 
	   	\label{eq:352¥}
   \end{eqnarray}¥
The boundary size $s_1$ between these sums is determined as follows. As schematically 
illustrated in Fig.\ref{fig.055¥}, the wavelength $2\pi /q$ of the transverse excitation with $\omega =a_tq^2$ 
slightly exceeds  the length of the condensate $R_{\parallel}$ at $s_1$. Here, we apply the model 
calculation described  in Sec 2B to this condition, that is, 
in the ellipsoid with a volume $(4\pi/3)R_{\parallel }R_{\perp 
}^2=sr_0^3$ in which $R_{\parallel }=s^Zr_0$, 
this $s_1$ is determined by   $R_{\parallel }=2\pi /q$: that is,  
$s_1^Zr_0=2\pi /\sqrt{\omega /a_t}¥$, hence 
\begin{equation}
      s_1=\left(\frac{2\pi}{r_0¥}¥\sqrt{\frac{a_t}{\omega¥}¥}¥\right)^{1/Z} .
	   	\label{eq:353¥}
\end{equation}¥
On the right-hand side of Eq.(\ref{eq:352¥}), since $\hbar\omega\ll\epsilon _0$ 
 in experiments, we can ignore the second term.  In the first term, 
 at every $q$, a sum over  $s$ from $1$ to $s_1$ is calculated, 
 in such a way that the transverse excitation beyond the coherent wave 
 function has an energy of $a_tq^2$.  Putting Eq.(\ref{eq:352¥}) into 
 Eq.(\ref{eq:2999¥}), and integrating it over $q^2$ while  keeping $q^2/\omega$ fixed, 
 the summation  over $s$ using  $sh(s)=\exp (\beta\mu s)$ yields
  \begin{eqnarray}
 &a&\int dq^2 \sum_{s=1}^{s_1¥}¥sh(s)\frac{q^2}{\omega ¥}¥\delta(\omega -a_tq^2)
                               \nonumber\\  
&=& -\frac{a}{a_t^2¥}¥\sum_{s=1}^{s_1¥}\exp(\beta\mu s)
			 = -\frac{a}{a_t^2¥}f_B(0,T)[1-\exp (\beta\mu s_1)] .
			 	\nonumber\\  
	   	\label{eq:354¥}
   \end{eqnarray}¥
At $\omega =0$, since $s_1=\infty$ in Eq.(\ref{eq:353¥}), the transverse motion makes a 
finite contribution to $\sigma_1(\omega)$ as $ -(a/a_t^2¥)f_B(0,T)$ in 
Eq.(\ref{eq:354¥}), but does not produce a $\delta (\omega)$ peak.

\subsection {$\sigma_1(\omega)$ from the longitudinal motion  }
 A superfluid flow, which is the translational motion of a 
 macroscopic condensate, is the $q\rightarrow 0$ limit of the longitudinal 
 oscillation.  Originally, it does not directly depend on 
 the viscosity $\eta _n$. Rather, it is determined by the basic 
 parameters, although  it is indirectly related to $\eta _n$ 
 by the sum rule [Eq.(\ref{eq:2983¥})]. Hence, in view of the dimension of 
 $\sigma_{1}(\omega)$, and the experimental result that the flow rate 
 of a superfluid flow is inversely proportional to $d^2$, rather than 
 $L$, we assume the general form 
 $\widetilde {A}(T)(\pi/d^2¥) \delta (\omega)$, and 
 will determine a dimensionless quantity $\widetilde {A}(T)$ using the sum rule \cite {sup}. 
  On the other hand, for the longitudinal motion observed at $\omega \neq 0$ in 
 Eq.(\ref{eq:2999¥}),  there is no boundary size $s_1$ in the sum over $s$.  Hence,  we obtain
 \begin{eqnarray}
&a&\int dq^2\sum_{s=1}^{\infty¥}¥sh(s)\frac{q^2}{\omega ¥}¥\delta(\omega -a_lq^2) 
                    	\nonumber\\     
&=& -\frac{a}{a_l^2¥}¥\sum_{s=1}^{\infty¥}\exp(\beta\mu s)
    =  -\frac{a}{a_l^2¥}f_B(0,T).
	   	\label{eq:355¥}
   \end{eqnarray}¥

\subsection {Total $\sigma_1(\omega)$   }
Substituting Eqs.(\ref{eq:354¥}) and (\ref{eq:355¥}) into the 
integrand of Eq.(\ref{eq:2999¥}), we obtain
 \begin{eqnarray}
&a&\int dq^2 \sum_{s=1}^{\infty¥} ¥sh(s)¥\frac{q^2}{\omega¥}¥ 
                    \nonumber\\  
	&\times &	 \left[\delta(\omega-a_lq^2)-\delta(\omega-\epsilon_t(s,q)/\hbar)\right]¥ 
			      \nonumber\\       
&=&f_B(0,T) \left[-\frac{a}{a_l^2¥}¥+\frac{a}{a_t^2¥}¥-\frac{a}{a_t^2¥}¥\exp(\beta\mu s_1)\right]¥.
		      \label{eq:3585¥}
   \end{eqnarray}¥
   
Using Eq.(\ref{eq:3585¥}) in Eq.(\ref{eq:2999¥}), 
 we finally obtain the fluidity spectrum
 \begin{eqnarray}
&\sigma_{1}&(\omega) = \sigma_{1n}(\omega)
                  +  \widetilde {A}(T)\frac{\pi}{d^2¥}  \delta (\omega)
       + \frac{\hbar }{2m¥}¥\pi \frac{\rho _s(T)}{4\eta_n¥}¥ 
	            	\nonumber\\  
	 &\times&  \left[\frac{a }{a_t^2¥}¥-\frac{a }{a_l^2¥}-\frac{a}{a_t^2¥}¥
	  \exp \left(\beta\mu(T) 
	  \left[\frac{2\pi}{r_0¥}¥\sqrt{\frac{a_t}{\omega ¥}¥}¥\right]^{1/Z}¥\right)¥
	             +\frac{a}{a_l^2¥}¥¥\right]¥ 
				   \nonumber\\  
              	\label{eq:359¥}
\end{eqnarray}¥
where Eq.(\ref{eq:353¥}) is used for $s_1$, and $\rho _s(T)=mf_B(0,T)$ is used 
for the small but nonzero condensate already existing just above 
 $T_{\lambda}$ ($m$ is the conventional mass in thermodynamic quantities).
In this result, the last  $a/a_l^2$ is added in the bracket of the third term, in order 
to ensure $\sigma_{1}(\omega)\rightarrow 0$ at $\omega \rightarrow \infty$.
 Using $a=\hbar/(2m)$ and $\epsilon _0=\hbar a_t/r_0^2$ 
for the transverse motion, we rewrite Eq.(\ref{eq:359¥})  as 
 \begin{eqnarray}
\sigma_{1}(\omega) &=& \sigma_{1n}(\omega) 
                     + \widetilde {A}(T)\frac{\pi}{d^2¥}  \delta (\omega)
			       \nonumber\\
       &+& \pi\frac{m_t^2}{m^2}\frac{\rho _s(T)}{4\eta_n¥}¥ 
	                  \nonumber\\  
	   &\times&\left[1-\exp \left(\beta\mu(T)
	   \left[\sqrt{(2\pi)^2¥\frac{m}{m_t¥}¥\frac{\epsilon _0}{\hbar\omega ¥}¥}¥\right]^{1/Z}¥\right)¥\right]¥ .
	            \nonumber\\  
	         	\label{eq:360¥}
\end{eqnarray}¥
[For liquid helium 4, $(2\pi)^2\epsilon _0 /\hbar\omega \simeq 1.11 \times 10^{14}/\omega$.]

  Equation (\ref{eq:360¥}) is a  microscopic representation of the change from
$\sigma (\omega)$ to $\sigma_s(\omega)+A\delta (\omega)$ in Fig.\ref{fig.015¥}. 
If the statistical gap does not exist in the transverse excitation, 
only the first and second terms remain on the right-hand side of 
Eq.(\ref{eq:360¥}), which reduces to the simplest two-fluid model 
representing the  discontinuous appearance of a superfluid flow. 
Hence, the statistical gap $\epsilon_0$ is necessary 
 for the continuous growth  of a superfluid flow.
Although  $\epsilon _0$  itself does not explicitly appear in 
 $\widetilde {A}(T)(\pi/d^2¥)\delta (\omega)$, the statistical gap  
 controls the superfluid flow behind the scenes.

\subsection {$\sigma_{1n}(\omega)$ from normal-fluid component  }

 The two-fluid picture  gives us the fluid-mechanical 
equation by rewriting Eq.(\ref{eq:002¥}) with $\rho =\rho _n(T)+\rho _s(T)$ and
$\mbox{\boldmath $v$}=\mbox{\boldmath $v$}_n(T)+\mbox{\boldmath $v$}_s(T)$ as
\begin{equation}
 \rho _n\frac{\partial\mbox{\boldmath $v$}_n}{\partial t¥}¥+\rho _s\frac{\partial\mbox{\boldmath $v$}_s}{\partial t¥}¥
            = \eta¥ \nabla^2\mbox{\boldmath $v$}_n- \nabla P .
          \label{eq:2901¥}
\end{equation}¥
Since $\mbox{\boldmath $v$}_s$ is a potential flow as 
$\mbox{\boldmath $v$}_s=\nabla \phi$, where $\phi$ is a potential 
function, we obtain the equation of $\mbox{\boldmath $v$}_n$ as
\begin{equation}
 \rho _n\frac{\partial\mbox{\boldmath $v$}_n}{\partial t¥}¥
            = \eta¥ \nabla^2\mbox{\boldmath $v$}_n- \nabla P_n ,
          \label{eq:2902¥}
\end{equation}¥
where $P_n$ is the pressure acting on the normal-fluid component 
\begin{equation}
 P_n=P+\rho _s(T)\frac{\partial\phi}{\partial t¥}¥ .
          \label{eq:2903¥}
\end{equation}¥
 Following the Newtonian equation of motion,  the superfluid flow is 
 accelerated under pressure $P_0$ at $z=0$ as
\begin{equation}
v_s=\left(\frac{P_0\pi d^2}{m¥}¥\right)¥t ,
          \label{eq:2904¥}
\end{equation}¥
and therefore the potential function in $\mbox{\boldmath 
$v$}_s=\nabla \phi$  has the form 
\begin{equation}
\phi (z,t)=\left(\frac{P_0\pi d^2}{m¥}¥\right)¥tz .
          \label{eq:2905¥}
\end{equation}¥
 Instead of the original pressure gradient  $P/L$ given by 
 $P(z)=P_0(1-bz)$, by using Eq.(\ref{eq:2905¥}) in Eq.(\ref{eq:2903¥}), 
 we find that the normal-fluid component in liquid experiences a 
 different pressure of
\begin{equation}
P_n(z)=P_0\left(1-\left[b-\rho _s(T)\frac{\pi d^2 }{m¥}¥\right]z¥\right)¥ .
          \label{eq:2906¥}
\end{equation}¥
In Eq.(\ref{eq:2902¥}), the pressure gradient takes a form such as
\begin{equation}
\nabla P_n=\left[1-\rho _s(T)\left(\frac{\pi d^2 }{bm¥}¥\right)¥\right]¥\nabla P
            \equiv [1-\rho _s(T)f]\nabla P.
          \label{eq:2907¥}
\end{equation}¥
The mass transport by the superfluid flow reduces the pressure 
gradient acting on the normal-fluid component. Hence, as a 
solution of Eq.(\ref{eq:2902¥}), the fluidity spectrum 
$\sigma _{1n}(\omega)$ of the normal-fluid component is given by
 \begin{eqnarray}
  \sigma _{1n}(\omega) =&-&\left(\frac{1-\rho _s(T)f}{\omega d^2¥}\right)¥
	                       \nonumber\\
	&\times&   Im\left(1-\frac{1}{J_0\left(id[1-i]\displaystyle {\sqrt {\frac{\rho_n(T)}{2\eta _n¥}¥\omega}}\right)¥¥}\right)¥.  
           \label{eq:2908¥}
\end{eqnarray}¥

\subsection {Sum rule  }

 The change in $\sigma_{1}(\omega)$ near $T_{\lambda}$ must satisfy 
 the sum rule.  Using Eq.(\ref{eq:293¥}) for $\sigma (\omega)$ on the 
 left-hand side of Eq.(\ref{eq:2983¥}), and using Eqs.(\ref{eq:360¥}) 
 and (\ref{eq:2908¥}) for $\sigma_s(\omega) +A\delta(\omega) $
 on the right-hand side, we obtain the sum rule
 \begin{eqnarray}
  \widetilde {A}(T)\frac{\pi}{d^2¥}
     &+&\pi\frac{\rho _s(T)}{4\eta_n¥}¥ 2\frac{m_t^2}{m^2¥}\int_{0}^{\infty¥}¥d\omega
			 \nonumber\\	 
	 &\times & \left[1-\exp \left(\beta\mu(T)
 \left[\sqrt{(2\pi)^2¥\frac{m}{m_t¥}¥\frac{\epsilon _0}{\hbar\omega ¥}¥}¥\right]^{1/Z} ¥\right)¥¥\right]¥
	                       \nonumber\\
	  &=&2\int_{0}^{\omega _c¥}¥ \frac{d\omega}{\omega d^2¥}
	         Im\left(\frac{1}{J_0\left(id[1-i]\displaystyle {\sqrt 
			 {\frac{\rho}{2\eta _n¥}¥\omega}}\right)¥¥}\right)¥
			              \nonumber\\
     &-&2\int_{0}^{\omega _c¥}¥ \frac{d\omega}{\omega d^2¥}
		       Im\left(\frac{1-f\rho _s(T)}{J_0\left(id[1-i]\displaystyle {\sqrt 
		  {\frac{\rho_n(T)}{2\eta _n¥}¥\omega}}\right)¥¥}\right)¥ ,
		             \nonumber\\
              	\label{eq:365¥}
\end{eqnarray}¥
where $\rho_n(T)+\rho_s(T)=\rho$.  Whereas the right-hand side of 
Eq.(\ref{eq:365¥}) comes from fluid mechanics including 
phenomenological parameters, the left-hand side comes from the 
microscopic many-body theory. 
This sum rule for a liquid can be regarded as a dynamic version of 
the equation of states  of the ideal Bose gas
\begin{equation}
  \frac{1}{V¥}¥\frac{z}{1-z¥}¥+\frac{1}{\lambda ^3¥}¥ g_{3/2}(z)=\frac{N}{V¥} 
	   	\label{eq:3742¥}
\end{equation}¥
[$z=\exp (\beta\mu)$, and $\lambda$ is the thermal wavelength]. 
Both equations determine the static or dynamic representation of 
a superfluid by the conservation law.

Historically, the striking finding of superfluidity was a discovery 
of a frictionless flow with $\eta =0$. However, its more essential 
feature is not $\eta =0$, but the constraint $rot \mbox{\boldmath $v$}_s=0$ 
on the superfluid flow \cite {lon}. Equation (\ref{eq:365¥}) supports 
this view as follows. The existence of the statistical gap $\epsilon 
_0$, which forbids transverse motion such as $rot \mbox{\boldmath $v$}_s=0$, 
induces the change in $\sigma_{1}(\omega)$.  As a result,  
$\widetilde {A}(T)(\pi/d^2)\delta (\omega)$ representing the 
frictionless flow $\eta =0$ inevitably appears so as to satisfy the sum rule.

(1) For the right-hand side of Eq.(\ref{eq:365¥}),  fluid mechanics does not describe 
oscillations with high frequency, because a liquid shows solidlike 
properties at a  high frequency \cite {rec}. 
Hence,  we cannot extend the upper end of the integral to $\infty$
and assume a cutoff frequency  $\omega _c$.

(2) The effect of a repulsive interaction on the superfluid flow is unclear. Hence, 
in the model calculation in Sec 5, we will assume no renormalization of the mass
$m_t=m_l=m$ as a first approximation, and  determine $\tilde {A}(T)$ of the superfluid flow so that 
Eq.(\ref{eq:365¥}) is satisfied at each temperature.

\section  {Model Calculation of Fluidity Spectrum  }

Experimentally, we observe a seemingly vertical drop of  $\nu(T)=1/(4\sigma(T))$ 
 at $T_{\lambda}$ as in Fig.\ref{fig.005¥}. If we could  put our 
 thought experiment into practice under the  precise control of  
 temperature and suppress the fluctuation near $T_{\lambda}$,
 we could observe the growth of a superfluid flow as a change in the 
 fluidity spectrum $\sigma_{1}(\omega)$.  All the results in Sec.5 
 are obtained so that they satisfy the sum rule in Eq.(\ref{eq:365¥}).

 In comparison with a gas, a strongly coupled system such as a liquid has more excitations with low energy. 
Hence, the temperature dependence of $\mu (T)$ in a liquid is weaker than $\mu _0(T)$ in a  gas.
However, the chemical potential $\mu (T)$ of liquid helium 4 just 
above $T_{\lambda}$ has not yet derived from experiment and theory.
When we use $\mu _0(T)$ [Eq.(\ref{eq:055¥})] of the ideal 
Bose gas  as $\mu (T)$ in $h(s)$, such a $h(s)$ extends to a large $s$  
 only after $T$ becomes extremely close to $2.17K$.
Because of these  reasons, ambiguity remains in our model calculation of the
fluidity spectrum for liquid helium 4 \cite {pre}. 
Rather, to demonstrate the mechanism in Fig.\ref{fig.055¥} without ambiguity, 
we perform a model calculation of it at extremely close $T$ to $2.17K$.

 We are interested in the effect of the repulsive 
interaction $U_0$ on the superfluid flow through different $\rho _s(T)/\rho$.
Numerically, we adopt a definition $\rho_s(T)/\rho=f_B(0,T)/\widehat {f}_B(0,T_{\lambda})$
in which $U_0$ in $f_B(0,T)$ and $U_0'$ in $\widehat {f}_B(0,T)$ are 
different. By changing $U_0/U_0'$, we control $\rho _s(T)/\rho$.
We are also interested in the effect of the degree of elongation 
 $Z$ of the coherent wave function on the fluidity spectrum. 
 We assumed an elongated form of the coherent wave function in the flow 
 as $R_{\parallel }\simeq s^Z r_0$.  In summary, we consider three cases: 
 (a) $\rho _s(T_{\lambda})/\rho =1.00$ and $Z=1$,  (b) $\rho 
 _s(T_{\lambda})/\rho =0.56$ and $Z=1$, and (c)  $\rho 
 _s(T_{\lambda})/\rho =1.00$ and $Z=0.9$,  by assuming
$U_0/U_0'=1.0$ for (a)  and (c) and $U_0/U_0'=1.8$  for (b).

Other parameters used are as follows. In the normal phase of liquid helium 4, we have 
$\rho=1.43\times 10^{2}$ [kg/$m^{3}$] and $\eta _n=3.0\times 10^{-6}$ 
[$Js/m^{3}$]: hence, $\rho/(4\eta _n) =1.2\times 10^7$ [$s/m^2$]. 
In Eq.(\ref{eq:365¥}), $d=1mm$, $\omega _c=0.6s^{-1}$ and 
$f=\pi d^2/(bm)=0.1$ are used.

 \begin{figure}
		\begin{center}
\includegraphics [scale=0.68]{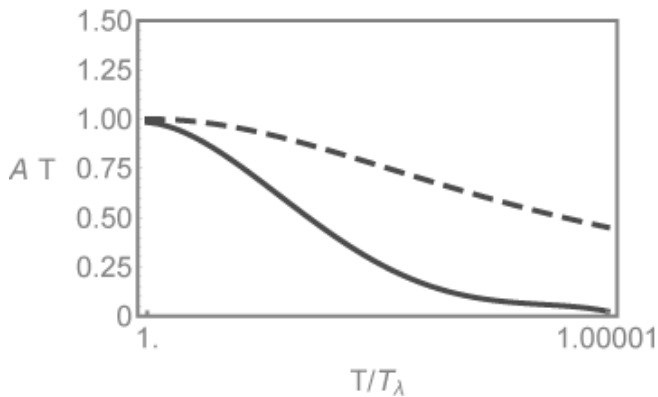}
\caption{\label{fig:epsart} Growth of superfluid flow $\widetilde 
{A}(T)$ (thick curve) just above $T_{\lambda}$ for 
(a) $\rho _s(T_{\lambda})/\rho =1.00$ and $Z=1$.  The dashed
curve shows $\rho _s(T)/\rho$.  }
\label{fig.047¥}
     	\end{center}
\end{figure}

\subsection   {Growth of superfluid flow just above $T_{\lambda}$}

 \begin{figure}
		\begin{center}
\includegraphics [scale=0.68]{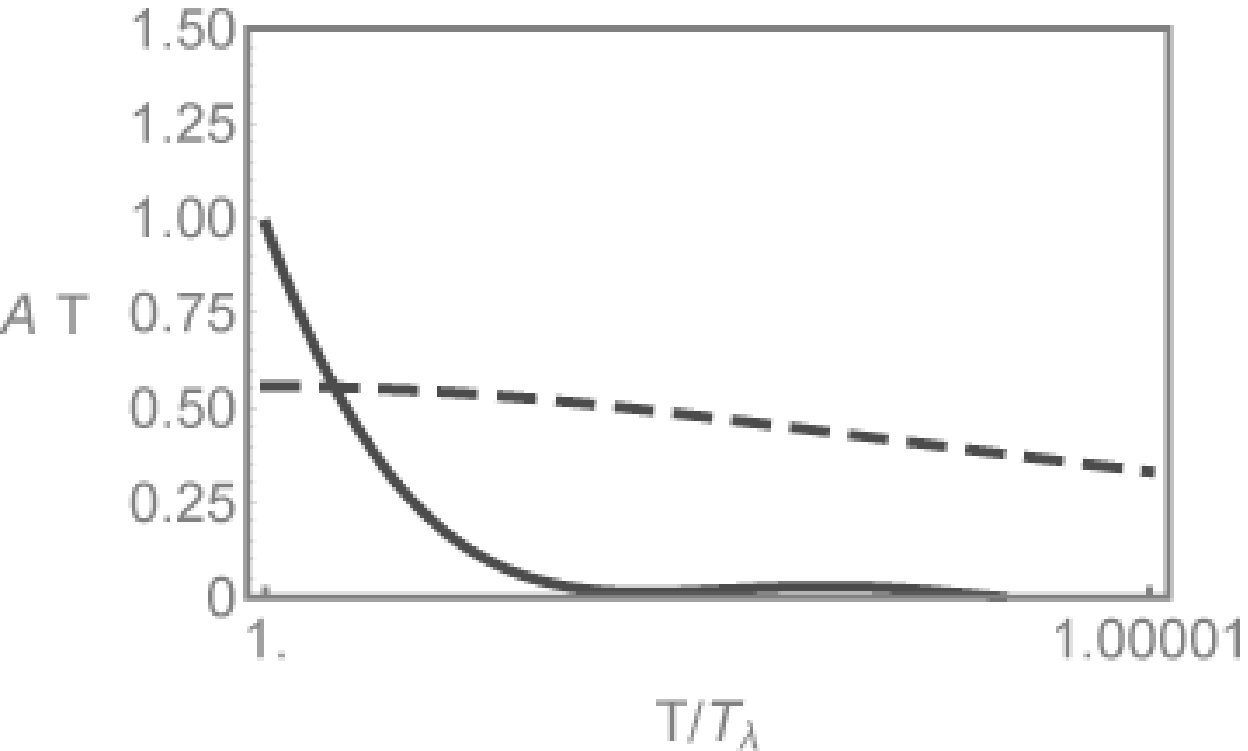}
\caption{\label{fig:epsart} Growth of superfluid flow $\widetilde 
{A}(T)$ (thick curve) just above $T_{\lambda}$ for
(b) $\rho _s(T_{\lambda})/\rho =0.56$ and $Z=1$. The dashed 
curve shows $\rho _s(T)/\rho$.  }
\label{fig.0471¥}
     	\end{center}
\end{figure}

 \begin{figure}
		\begin{center}
\includegraphics [scale=0.68]{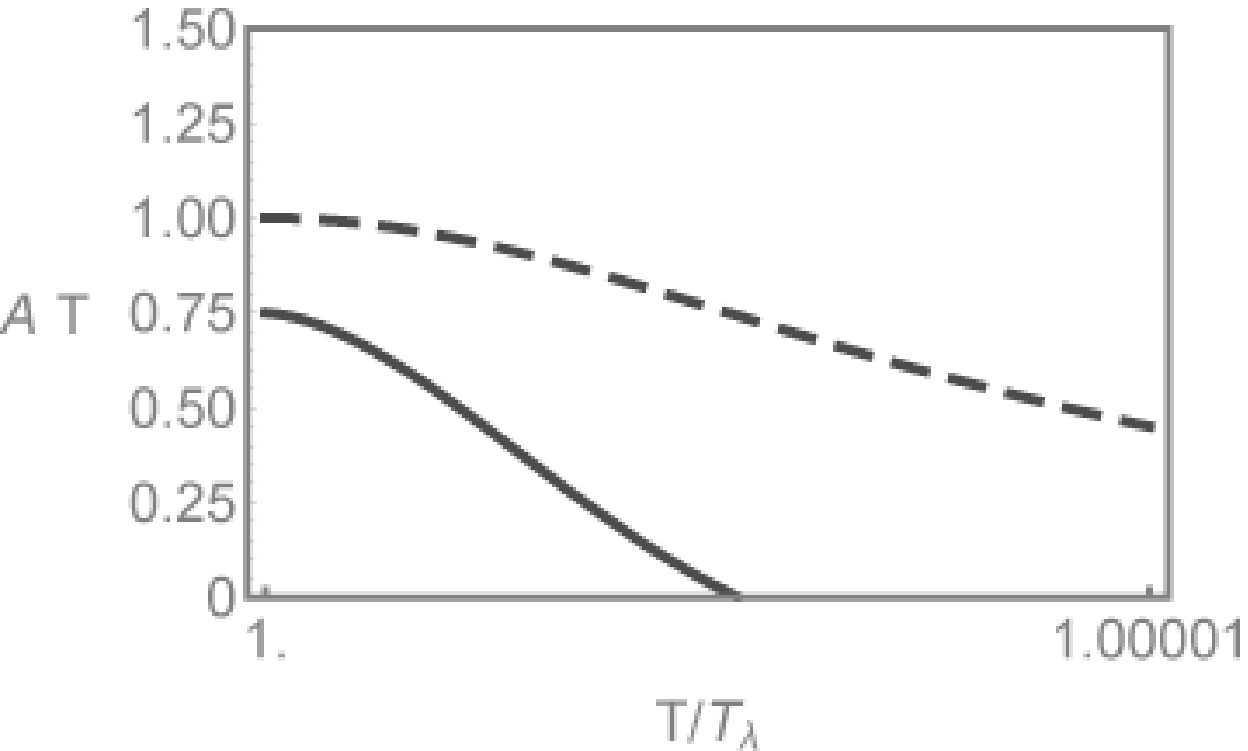}
\caption{\label{fig:epsart} Growth of superfluid flow $\widetilde 
{A}(T)$ (thick curve) just above $T_{\lambda}$ for (c)  
$\rho _s(T_{\lambda})/\rho =1.00$ and $Z=0.9$. The dashed 
curve shows $\rho _s(T)/\rho$.  }
\label{fig.0472¥}
     	\end{center}
\end{figure}

The rate of the superfluid flow  $\widetilde {A}(T)$ in $\widetilde {A}(\pi/d^2¥)¥\delta(\omega) $
just above $T_{\lambda}$ is determined as 
a function of $T/T_{\lambda}$ so as to satisfy the sum rule  Eq.(\ref{eq:365¥}). 
 Figure \ref{fig.047¥} shows  $\widetilde {A}(T)$ just above 
 $T_{\lambda}$ (thick curve) for (a)  $\rho _s(T_{\lambda})/\rho =1.00$ and $Z=1$, 
 in comparison with $\rho _s(T)/\rho$ (dashed curve). 
  Figure \ref{fig.0471¥} shows  $\widetilde {A}(T)$ just above 
 $T_{\lambda}$  for (b)  $\rho _s(T_{\lambda})/\rho =0.56$ and $Z=1$, 
 in comparison with  $\rho _s(T)/\rho$. 
  For both (a) and (b), $\widetilde {A}(T)$ grows  more rapidly 
 than $\rho _s(T)/\rho$ near $T_{\lambda}$. In the case of (b) for $\rho 
 _s(T_{\lambda})/\rho =0.56$,  $\rho _s(T)/\rho $ remains smaller than $1$, but 
  $\widetilde {A}(T)$  finally rises to $1$ at $T_{\lambda}$. In view 
  of these results, the thermodynamic  rate of the superfluid density 
  $\rho _s(T)/\rho$ and the rate of the translational superfluid 
  flow $\widetilde {A}(T)$ are different quantities.

Figure \ref{fig.0472¥} shows $\widetilde {A}(T)$ and $\rho _s(T)/\rho$ for 
(c) $\rho _s(T_{\lambda})/\rho =1.00$ and $Z=0.9$. 
If we assume a more isotropic form of the coherent wave function  in 
the flow by reducing $Z$ in $s^Zr_0$, the effective length $R_{\parallel }$ of 
the coherent wave function in Fig.\ref{fig.055¥} is 
 reduced in comparison with the case of $Z=1$ at the same temperature. 
 Hence, $\widetilde {A}(T)$ does not reach $1$ at $T_{\lambda}$.

\subsection   {Precursory form of superfluid flow in the increase in $\sigma_{1}(\omega)$ near $\omega =0$ }

\begin{figure}
		\begin{center}
\includegraphics [scale=0.65]{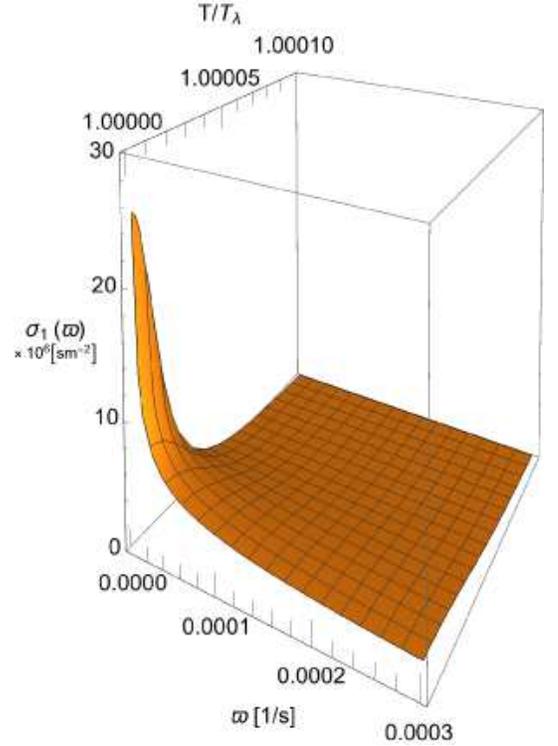}
\caption{\label{fig:epsart} Continuous increase in the fluidity spectrum 
$\sigma _1(\omega)¥$ near $\omega =0$ along the axis of capillary 
when approaching $T_{\lambda}$  in the oscillatory  Poiseuille's flow.   
The third term in Eq.(\ref {eq:360¥})  is plotted for (a)  $\rho 
_s(T_{\lambda})/\rho =1.00$ and $Z=1$. [$\sigma_{1n}(\omega)$ is not included.] } 
\label{fig.05¥}
     	\end{center}
\end{figure}

  Using the third term on the right-hand side of Eq.(\ref {eq:360¥}), 
$\sigma_{1}(\omega)$ near $\omega =0$ is plotted. 
 Figure \ref{fig.05¥} shows the continuous increase in $\sigma_{1}(\omega)$ near $\omega =0$
 for (a) $\rho _s(T_{\lambda})/\rho =1.00$ and $Z=1$ just above $T_{\lambda}$.
 [$\sigma_{1n}(\omega)$ is not included.]
 This temperature dependence comes from $\mu (T)$ in the exponent  and $\rho _s(T)$. 
  At $1.00003 T_{\lambda}$,  the 
 peak at $\omega =0$ has not yet appeared, because only the small 
 coherent wave function contributes to $\sigma_{1}(\omega) $. At 
 $1.00002 T_{\lambda}$, a small peak at $\omega =0$  appears.
 As $T\rightarrow T_{\lambda}$, the effect of the statistical gap
 gradually accumulates in  $\sigma_{1}(\omega) $  at low frequencies, finally 
 becoming the $\delta (\omega)$-function peak representing a superfluid flow.  
  The abrupt change in $\sigma_{1}(\omega)$ just above $T_{\lambda}$ 
in Fig.\ref{fig.05¥} is consistent with the observed vertical drop of $\nu$ 
 in Fig.\ref{fig.005¥}. 
  Since $s_1$ in Eq.(\ref{eq:353¥}), determining the $\omega$ dependence of 
 $\sigma_1(\omega)$ near $\omega =0$, does not depend on $\rho _s(T)/\rho$, 
the continuous increase in $\sigma_{1}(\omega)$ near $\omega =0$ for (b) 
just above $T_{\lambda}$  is  similar to that of (a). However, since $s_1$ depends on $Z$, 
 $\sigma_1(\omega)$ near $\omega =0$ for (c) $Z=0.9$ is considerably 
 different from those of (a) and (b).

 The increase in $\sigma_1(\omega)$ at small $\omega$ in Eq.(\ref{eq:2999¥}) is due to excitations with 
a large wavelength $\lambda$ for $q=2\pi/\lambda$ in $\delta (\omega -aq^2)$. 
Hence, in the sum over $s$ in  Eq.(\ref{eq:352¥}) at small $\omega$,  a difference
between the longitudinal and transverse excitations continue to exist 
from a small  to large $s$, thereby leading to a large $\sigma_1(\omega)$ near
$\omega =0$. In contrast,  $\sigma_1(\omega)$ at
large $\omega$ is due to  excitations with a small wavelength. 
Hence, the difference between both excitations exists only for  
 small $s$ in the sum, hence leading to a small 
 $\sigma_1(\omega)$.  Just above $T_{\lambda}$, the approach 
 of $\mu(T)$ to zero as $T\rightarrow T_{\lambda}$ in the exponent of Eq.(\ref{eq:360¥})
 represents the growth of the condensate,  hence leading to the inclusion of a long excitation  in 
$\sigma_{1}(\omega)$, as  schematically illustrated in Fig.\ref{fig.055¥}. 
The exponent in Eq.(\ref{eq:360¥}) shows this mechanism at different temperatures. 
The third term in Eq.(\ref{eq:360¥}) has a comparable value with the 
first term  $\sigma_{1n}(\omega) $ only at $\omega \simeq 0$. For larger $\omega$, 
it rapidly falls owing to the exponential factor, 
and only $\sigma_{1n}(\omega) $ remains in $\sigma_{n}(\omega) $.

\subsection   {$\sigma_{1n}(\omega)$ of normal-fluid component}

 \begin{figure}
		\begin{center}
\includegraphics [scale=0.68]{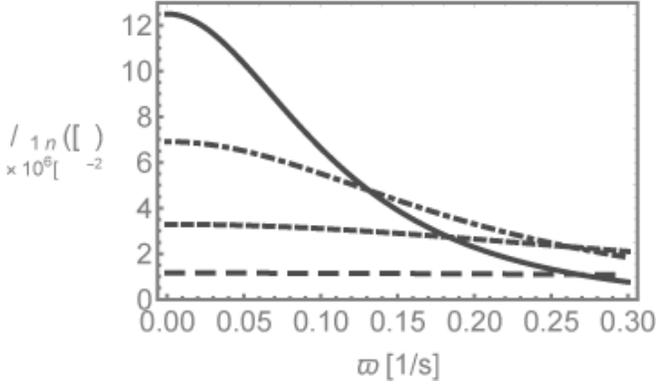}
\caption{\label{fig:epsart} Fluidity spectrum $\sigma 
_{1n}(\omega)¥$ of normal-fluid component 
just above $T_{\lambda}$ in the oscillatory  
Poiseuille's flow for (a). Equation (\ref{eq:2908¥}) is plotted 
 in the case of $f=0.1$ and $d=1 mm$, with $\rho _n$ at $T\gg T_{\lambda}$ (thick curve),  
$\rho _n$ at $1.00001T_{\lambda}$ (one-point dotted curve),  $\rho 
_n$ at $1.0000055T_{\lambda}$ (short dotted curve), and  $\rho _n$ 
at $1.000003T_{\lambda}$ (long dotted curve).   }
\label{fig.045¥}
     	\end{center}
\end{figure}

 When  the $\delta (\omega)$-function peak begins to rise when cooling the system,
$\sigma_{1}(\omega)$ is still composed of  $\sigma_{1n}(\omega)$ of the normal-fluid component. 
 This $\sigma_{1n}(\omega)$ in Eq.(\ref{eq:2908¥}) depends on not only $\rho/\eta$, 
but also the radius $d$ of the capillary. 
 Figure \ref{fig.045¥} shows the temperature change of 
 $\sigma_{1n}(\omega)$ in Eq.(\ref{eq:2908¥}) for (a) and (c) in the case of $f=0.1$ and 
$d=1$ [mm]: The $\sigma_{1n}(\omega)$  at $T\gg T_{\lambda}$ (thick curve) changes to  
$\sigma_{1n}(\omega)$ at $1.00001T_{\lambda}$  (one-point dotted curve) and    
$\sigma_{1n}(\omega)$ at $1.000005T_{\lambda}$ (short dotted curve), 
and finally reaches  $\sigma_{1n}(\omega)$  at $1.000003T_{\lambda}$ (long dotted curve).  
Throughout this change,  the upper-convex form of $\sigma_{1n}(\omega)$ near $\omega =0$ 
 is maintained, which is a characteristic feature of a normal fluid.

\subsection   {Total fluidity spectrum $\sigma_{1}(\omega)$ } 

\begin{figure}
		\begin{center}
\includegraphics [scale=0.68]{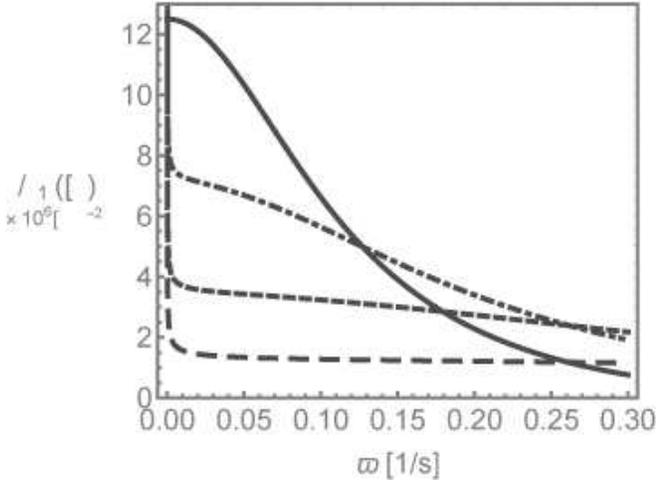}
\caption{\label{fig:epsart} Continuous change in the total fluidity 
spectrum $\sigma _1(\omega)¥$  when approaching $T_{\lambda}$. They 
are calculated using Eq.(\ref {eq:360¥}), in which $\tilde {A}(T)$ 
satisfies Eq.(\ref{eq:365¥}), for (a)  $\rho _s(T_{\lambda})/\rho 
=1.00$ and $Z=1$.
The thick upper-convex curve shows $\sigma_{1n}(\omega)$ 
 in the normal phase at $T \gg T_{\lambda}$ (Fig.\ref{fig.045¥}).
The one-point dotted curve appears at  
 $1.00001$ $T_{\lambda}$.  
 In the short dotted curve at  $1.0000055$ $T_{\lambda}$, 
 the upper-convex  nature near $\omega=0$ has already disappeared. 
At $1.000003$ $T_{\lambda}$,  just above $T_{\lambda}$,
this  $\sigma_{1}(\omega)$ changes to the long dotted curve.  }. 
\label{fig.07¥}
     	\end{center}
\end{figure}

 \begin{figure}
		\begin{center}
\includegraphics [scale=0.68]{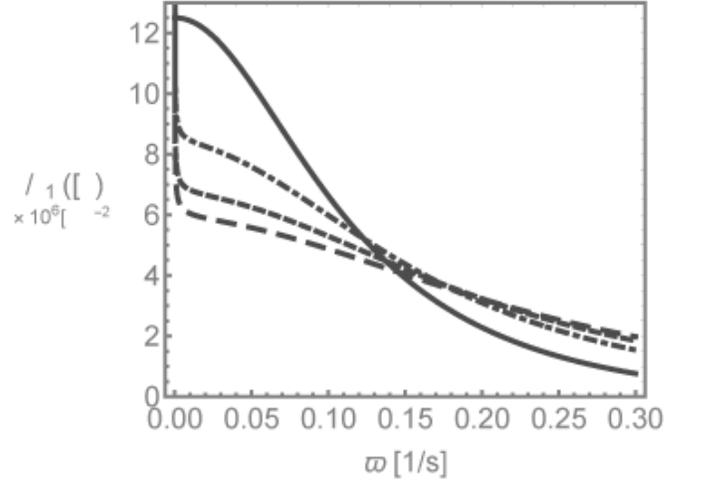}
\caption{\label{fig:epsart} Continuous change in the total fluidity 
spectrum $\sigma _1(\omega)¥$ when approaching $T_{\lambda}$, 
for (b)  $\rho _s(T_{\lambda})/\rho =0.56$ and $Z=1$.  }. 
\label{fig.071¥}
     	\end{center}
\end{figure}

 Finally,  we obtain a more precise form of $\sigma_s(\omega)+A\delta(\omega)$, 
 which was schematically illustrated in Fig.\ref{fig.015¥},  by 
 combining  the  microscopic and phenomenological methods.  

 Figure \ref{fig.07¥} shows the total fluidity spectrum $\sigma_{1}(\omega)$  
 in Eq.(\ref{eq:360¥}) for (a)  $\rho _s(T_{\lambda})/\rho =1.00$ and $Z=1$
 at four different temperatures just above $ T_{\lambda}$ 
 under the same parameters  used in Fig.\ref{fig.05¥}. 
 The thick upper-convex curve shows $\sigma_{1n}(\omega)$ 
 in the normal phase at $T \gg T_{\lambda}$ [Eq.(\ref{eq:293¥})] 
 (thick curve in Fig.\ref{fig.045¥}). 
 As $T\rightarrow T_{\lambda}$, $\sigma_{1}(\omega)$ continuously 
 changes as follows.  The one-point dotted curve at  
 $1.00001 T_{\lambda}$ has already lost the upper-convex form near 
 $\omega=0$,  which transforms to a lower-convex one.  
 Upon cooling to $1.0000055 T_{\lambda}$,
  the short dotted curve of  $\sigma_{1}(\omega)$ appears. Finally,
 the long dotted curve  of $\sigma_{1}(\omega) $ appears at 
 $1.000003T_{\lambda}$ just above $T_{\lambda}$. 
 Similarly, Fig.\ref{fig.071¥} shows the continuous change in the total fluidity 
 spectrum $\sigma_{1}(\omega)$ for (b)  $\rho _s(T_{\lambda})/\rho =0.56$ and $Z=1$,
 in which $\sigma_{1n}(\omega) $ of the normal-fluid 
 component still remains just above $T_{\lambda}$.

 The fluidity $\sigma_{1}(\omega)$ near $\omega =0$ changes its form
 from an upper-convex form to a lower-convex one.  The upper-convex form of 
  $\sigma_{1}(\omega)$ means that the classical liquid is robust when the
 applied  pressure slowly oscillates. This is because the viscous flow itself 
contains oscillations with various low frequencies.  Such an 
upper-convex form of $\sigma_{1}(\omega)$
 in the viscous flow continuously  changes to a lower-convex one in the 
 superfluid flow.  The lower-convex form means that the  
superfluid flow is very fragile when the applied  pressure oscillates. This is 
because the superfluid flow is an one-directional motion not including oscillations.

 Until now in this paper, we have been considering the thought 
 experiment in which an oscillatory 
 pressure is intentionally applied to a superfluid flow, but if it is 
 accidentally applied to a superfluid flow, we will have a different 
 impression on  the result.  When the flow velocity increases, 
 the friction against the wall of the capillary yields 
 the local oscillation of liquid, thus inducing pressure oscillation with 
a low frequency.  The lower-convex form of $\sigma _1(\omega)¥$ in 
 Figs.\ref{fig.07¥} and \ref{fig.071¥} indicates that even if the pressure only slightly 
 oscillates, it substantially destroys  the superfluid flow, hence 
 giving us a new interpretation of  the critical velocity $v_c$ of a superfluid flow
 at $T<T_{\lambda}$. This suppression is different from the well-known 
  mechanism of the critical superfluid velocity owing to the 
  formation of a vortex ring \cite {lan}.
  To predict a concrete value of $v_c$, however,  we must have a microscopic theory of the 
 friction arising between the wall of a capillary  and the flowing liquid.

\begin{figure}
		\begin{center}
\includegraphics [scale=0.68]{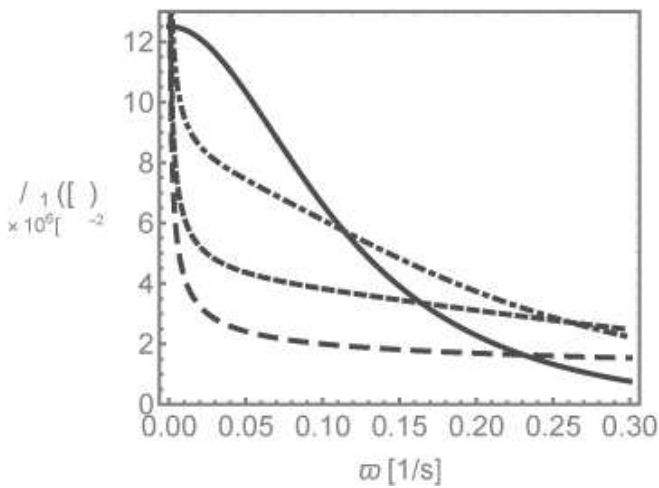}
\caption{\label{fig:epsart} Continuous change in the total fluidity 
spectrum $\sigma _1(\omega)¥$ when approaching $T_{\lambda}$, 
for (c) $\rho _s(T_{\lambda})/\rho =1.00$ and $Z=0.9$.  }. 
\label{fig.072¥}
     	\end{center}
\end{figure}

 Figure \ref{fig.072¥} shows the total fluidity spectrum 
 $\sigma_{1}(\omega)$ for (c) $Z=0.9$. 
When $Z=0.9$ in Eq.(\ref{eq:353¥}), the same $\omega$ corresponds to a larger $s_1$ 
compared with the case of $Z=1$. Hence, a large coherent wave 
function contributes to $\sigma_{1}(\omega)$ compared with the case of 
$Z=1$. The increase in $\sigma_{1}(\omega)$ is not localized to $\omega\simeq 0$. The 
transformation of $\sigma_{1}(\omega)$ to $\sigma_{s}(\omega)+A\delta (\omega)$ 
comes not only from $\sigma_{1n}(\omega)$ in Fig.\ref{fig.045¥}, but 
also from the third term in Eq.(\ref{eq:360¥}). 
 
In response to the slowly varying external perturbation, 
the strongly interacting particles with high density,  
such as those in a liquid,  reaches a local equilibrium at each instant.
Hence, the system is described by a small number of local equilibrium 
variables, and the fluid-mechanical  description becomes a useful 
tool for such a system. Compared with this general trend, the 
reason why the Bose condensate, although being a 
macroscopic system,  has only a small number of variables 
is found in a unique mechanism.  By suppressing other excitations, Bose statistics
enhances the role of one-particle excitations from the condensate, 
which also  has only a small number of variables.  
 In the transformation of the upper-convex $\sigma_{1}(\omega )$ to 
 the lower-convex $\sigma_{1}(\omega )$ in Figs.\ref{fig.07¥}--\ref{fig.072¥}, 
 we see a transition from one mechanism to another 
in  reducing the number of variables of the macroscopic systems.

\section{Discussion}

 The following intuitive explanation for the result of this paper is possible. 
 In analogy to the shear transformation of 
solids,  Maxwell related the shear viscosity $\eta$ of a liquid to the 
relaxation time $\tau$ of the dissipative process in the liquid by 
$\eta=G\tau$, where  $G$ is the modulus of rigidity of the liquid
(Maxwell's relation \cite{max}). The relaxation time $\tau$ is related 
to the energy  difference $E-E_0$ before and after the relaxation as
$\hbar /\tau \propto |E-E_0|$. Hence,  the statistical gap
 in $|E-E_0|$ leads to the decrease in $\tau$, then leading to the decrease in $\eta$.  
In  this paper, we formulated this intuitive 
explanation in the framework of statistical mechanics.

\subsection{Comparison with superconductivity}

In superconductivity, a  sum rule similar to Eq.(\ref{eq:2983¥}) 
exists for the conductivity of metals (the  
oscillator-strength sum rule)  \cite {sup}. 
In this  sum rule, the appearance of the energy gap of Cooper pairs 
induces a missing area near $\omega =0$ in the integrand of the sum rule, hence 
leading to the $\delta (\omega)$ peak of the supercurrent.   
The energy gap of Cooper pairs plays two roles, (1) the bosonization of 
two fermions and (2) the induction of a supercurrent. Such an energy gap does not exist
in intrinsic bosons. Rather, the transformation of the fluidity 
spectrum from the upper-convex to the lower-convex form such as Fig.\ref{fig.015¥} 
results in a missing area in the  integrand,  hence 
leading to $A\delta(\omega)$ of a superfluid flow. Since this transformation is 
induced by the statistical gap, it controls the phenomenon behind  the scenes. 

Behind this difference, there is an important difference in the 
nature of the transition. The phenomenon occurring in fermions in the 
vicinity of $T_c$ is not gradual growth of the macroscopic coherent wave 
function, but the abrupt formation of Cooper pairs by the various 
combinations of two fermions. Once 
Cooper pairs form, they are already composite bosons  with a high 
density at a low temperature, and they immediately form the superfluid 
state. Hence, the $\delta (\omega)$ peak and the energy-gap structure of Cooper 
pairs are not smoothly  connected in the excitation spectrum of superconductors.

In contrast, helium 4 atoms at $T\gg T_{\lambda} $ are intrinsic bosons, 
although they have not yet formed a macroscopic coherent wave function. As 
$T \rightarrow T_{\lambda}$, the coherent wave function gradually 
grows. Hence, even if it occurs within a very narrow temperature 
region above $T_{\lambda}$, we can consider the growth process of a
superfluid flow as a transformation of the nonequilibrium steady process. 
In this case,  the $\delta (\omega)$ peak and  
$\sigma_{s}(\omega ) $ at $\omega \neq 0$ are smoothly connected in 
the spectrum, which reflects the growth process of a superfluid flow.

\subsection{Role of the repulsive interaction}

The role of the repulsive interaction in a superfluid flow is an important problem. 

(1) In general, the repulsive interaction stabilizes the condensate by 
preventing the fragmentation of the condensate \cite {noz2}. If it 
fragments into two condensates, the exchange energy due to the repulsive 
interaction makes it costly. Hence, the repulsive interaction is a 
premise of our thermodynamic consideration, and the size distribution 
$h(s)=\exp (\beta\mu s)/s$ of the coherent wave function is not an exception.

(2) The mean-field effect $U_0$ of the repulsive interaction 
considerably changes the fluidity spectrum as shown in Figs.\ref{fig.07¥} and \ref{fig.071¥}.
 As shown in Figs.\ref{fig.047¥} and \ref{fig.0471¥}, however,  
 $\widetilde {A}(T)$ of a superfluid flow does not depend on $U_0$ at $T_{\lambda}$.

(3)  The many-body effect of the repulsive interaction $U$ on a superfluid flow is a difficult  problem. 
To describe its many-body effect, the perturbation expansions of $\chi^{L}(\mbox{\boldmath $q$},\omega)$ 
and $\chi^{T}(\mbox{\boldmath $q$},\omega)$  in powers of $U$ are 
needed. Their structures are different, because the roles of the repulsive interaction in the 
longitudinal and transverse cases are considerably different.
  
As a first approximation, we here consider  only a chain reaction 
of the longitudinal excitation from the condensate, and the decay to it. 
In $\chi_{\mu\nu}(\mbox{\boldmath $q$},\omega )$ of Eq.(\ref{eq:2995¥}), 
we replace $\chi^L(\mbox{\boldmath $q$},\omega)$ at $\omega>0$ 
by a sum of bubble-chain-type diagrams as follows  
 \begin{eqnarray}
 \chi^L(\mbox{\boldmath $q$},\omega)
     &=& -\frac{\hbar^2}{4¥}¥ \sum_{s}^{\infty¥}
	 \left(\frac{sh(s)}{\hbar\omega-\hbar 
	 a_lq^2¥}¥+\frac{-sh(s)}{\hbar\omega+\hbar a_lq^2¥}¥\right)¥q^2
			    \nonumber \\
 &\times& \left[\sum_{n=0}^{\infty¥}¥U^n\left(\frac{sh(s)}{\hbar\omega-\hbar 
	 a_lq^2¥}¥+\frac{-sh(s)}{\hbar\omega+\hbar a_lq^2¥}\right)^n¥\right]¥.
	                  \nonumber \\
                   \label{eq:370¥}	
\end{eqnarray}¥

Substituting this $\chi^L(\mbox{\boldmath $q$},\omega)$ into 
$\chi^L(\mbox{\boldmath $q$},\xi)$ ($-\infty <\xi<\infty$) of Eq.(\ref{eq:29¥}),
we obtain a new fluidity spectrum in place of Eq.(\ref{eq:2997¥}),
 \begin{eqnarray}
\sigma_{1}(\omega) &=& \sigma_{1n}(\omega) +
                \frac{\hbar}{2\pi¥}\frac{1}{4\eta_n¥}¥a\int dq^2
		 \sum_{s}^{\infty¥}¥sh(s)¥ \frac{q^2}{\omega¥}¥ \int_{-\infty}^{\infty¥}d\xi 
		                    \nonumber \\
          &\times &   \frac{1}{\xi-\omega¥}¥
			\left(\frac{1}{\xi-a_lq^2¥}-\frac{1}{\xi-\epsilon_t(s,q)¥/\hbar}¥\right)¥
			    \nonumber \\
		&+& \frac{\hbar}{2\pi¥}\frac{1}{4\eta_n¥}¥a\int dq^2
		  \sum_{n=1}^{\infty¥}¥\frac{U^n}{\hbar^n¥}¥
		                       \nonumber \\
		  &\times &  \sum_{s}^{\infty¥}¥
		                [sh(s)]^{n+1}¥\frac{q^2}{\omega¥}¥ \int_{-\infty}^{\infty¥}d\xi 
            \frac{1}{\xi-\omega¥}¥
			\left(\frac{1}{\xi-a_lq^2¥}\right)^{n+1}¥.	
			                   \nonumber \\
	                      \label{eq:372¥}
\end{eqnarray}¥
 For the third term, after defining a new variable $x=\xi-a_lq^2$, the Hilbert 
 transformation of $1/x^n$ is performed. Such a formula is obtained by repeating 
 the differentiation of $1/(x+i\epsilon)=1/x-i\pi\delta(x)$ with respect to $x$  
 by  $n$ times,
\begin{equation}
  \frac{1}{\pi¥}¥\int_{-\infty¥}^{\infty¥}¥¥\frac{dx}{x-\omega¥}¥
              \frac{1}{x^n¥} =-\pi \frac{(-1)^n}{\Gamma¥(n)}¥¥
			  \delta ^{(n-1)}(\omega)  ,
	   	\label{eq:374¥}
\end{equation}¥
where $\delta ^{(n)}(\omega)$ is the $n$th derivative of $\delta (\omega)$. 
Applying this formula to Eq.(\ref{eq:372¥})  yields
 \begin{eqnarray}
\sigma_{1}(\omega) &=& \sigma_{1n}(\omega) 
        +\frac{\pi\hbar}{2¥}\frac{1}{4\eta_n¥}¥	a\int dq^2 
		                  \nonumber \\
	 &\times & 	\sum_{s}^{\infty¥}¥sh(s)¥\frac{q^2}{\omega¥}¥ 
			 \left[\delta(\omega-a_lq^2)-\delta(\omega-\epsilon_t(s,q)/\hbar)\right]¥
			    \nonumber \\
		&+& \frac{\pi\hbar}{2¥}\frac{1}{4\eta_n¥}¥a\int dq^2
		  \sum_{n=1}^{\infty¥}¥\frac{U^n}{\hbar^n¥}¥
		                     \nonumber \\
		 &\times & \sum_{s}^{\infty¥}¥[sh(s)]^{n+1}¥
		             \frac{q^2}{\omega¥}¥\frac{(-1)^{n+1}}{\Gamma¥(n+1)}¥¥ \delta ^{(n)}(\omega -a_lq^2)  
					          \nonumber \\
	                      	\label{eq:376¥}
\end{eqnarray}¥
The third term shows that more singular functions $\delta ^{(n)}(\omega)$ 
than $\delta (\omega)$ appear. 
However, the presence of $(-1)^{n+1}$ means that these functions almost cancel each 
other. Hence, it is unclear whether the many-body effect of the repulsive interaction 
enhances or suppresses the superfluid flow, at least within the bubble-chain approximation. 
For this reason, we used Eq.(\ref{eq:360¥}) in the sum rule.

For rotational superfluidity, the many-body effect of the repulsive interaction $U$ 
enhances  the anomalous rotational behavior of the Bose systems just above $T_{\lambda}$  
\cite {busk} \cite {mei}. For translational superfluidity, however, 
the many-body effect of $U$ does not  necessarily enhance the superfluid flow.
 The anisotropic effective masses $m_t$ and $m_l$ in the flow are 
 created as the many-body effect of the repulsive interaction.   
If $m_t$ increases in the sum rule [Eq.(\ref{eq:365¥})], the 
superfluid flow $ \widetilde {A}(T) $ will be 
suppressed. For more precise treatment, it is necessary to begin with the dynamical model of 
a liquid, which still remains as an undeveloped area.  
 
(4) The experiment using a capillary shows that the flow rate of the
superfluid flow does not depend on the length $L$ of the capillary,  
but strongly depends on its diameter. In this work, we focus on 
the flow velocity along the axis of the capillary. To derive the dependence 
of the superfluid flow velocity on the diameter, we must extend the 
formalism to any distance $r$ from the axis, and integrate these results 
over the distance from zero to the radius $d$ of the capillary.

\subsection{Gradual decrease in viscosity at $T_{\lambda}<T<3.7$ K}

In this paper, we focus on the abrupt decrease in the kinematic 
viscosity $\nu (T)$ at $T_{\lambda}$, but  
the  gradual fall of  $\nu (T)$ at $T_{\lambda}<T<3.7$ K in 
Fig.\ref{fig.005¥} is not yet explained.

 In the simplest two-fluid model, Eq.(\ref {eq:360¥})  gives 
\begin{equation}
    \sigma_{1}(\omega ) =\sigma_{1n}(\omega )+
			  \widetilde {A}(T)  \frac{\pi }{d^2¥}¥\delta (\omega ).
	   	\label{eq:3558¥}
\end{equation}¥
For comparison with the experiment, we define a frequency-averaged fluidity 
$\sigma=\int_{-\infty ¥}^{\infty ¥}¥\sigma (\omega)d\omega$ using this
$\sigma_1(\omega)$, obtaining the coefficient of kinematic viscosity $\nu (T)=1/(4\sigma)$ as 
\begin{equation}
  \nu (T)=¥\frac{\nu _n}{1+\displaystyle{ \frac{\widetilde {A}(T) }{\sigma_{1n}}¥ 
             \frac{\pi}{d^2¥}¥}¥¥}¥, 
		\label{eq:362¥}
\end{equation}¥
where $\nu _n$ ($=1/[4\sigma _{1n}]$) is $\nu$ of the normal phase.
This formula means that the gradual fall of viscosity at $T_{\lambda}<T<3.7$ K
is caused by the gradual increase in $\widetilde {A}(T)$ in Eq.(\ref{eq:362¥}).
 The gradual fall of $\nu (T)$ already begins at $2.8$ K$<T<3.7$ K, but $\widetilde {A}(T)$
is so small that it cannot explain such a large fall of $\nu (T)$.

Historically, this gradual fall has been tacitly attributed to thermal 
fluctuation, which has not been explicitly distinguished from the 
well-known anomalies at $|T/T_{\lambda}-1|<10^{-2}$. 
(The exception was Ref.5, in which the authors pointed out
that the fall of $\eta $ in the range of $T_{\lambda}<T<3.7$ K 
does not come from thermal fluctuations, but from the gradual
change in the particle distribution occurring above $T_{\lambda}$.)
If we adhere to the naive analogy to the ideal Bose gas, nothing occurs above
$T_{\lambda}$ except for the fluctuation. In view of the complex nature of a liquid,
however, it is unnatural to assume that the ideal Bose gas picture 
grasps the specific features of a strongly interacting dense system, 
even if it is liquid helium 4.

Physically, the interpretation that $\nu (T)$ in the range of $T_{\lambda}<T<3.7$ K 
comes from thermal fluctuation has the following difficulty.  
  The temperature width of 1.5 K between $T_{\lambda}$ and 3.7 K is  
too large for the thermal fluctuation to occur.  The temperature 
width $\delta T$ of the fluctuation is given by the well-known formula 
\begin{equation}
 \langle (\delta T)^2\rangle = \frac{k_BT^2}{C_V¥}¥, 
		\label{eq:50¥}
\end{equation}¥ 
in which $C_V$ is the specific heat of a small fluctuating  region \cite {stat}. 
 $C_V$ derived from $\delta T=1.5$ K in Eq.(\ref{eq:50¥}) is about
$2.8 \times 10^{-23}$ J/K. As mentioned in Ref.5, if the 
interpretation on the basis of fluctuation was true,  
 the fluctuating region corresponding to this value of $C_V$ would be as 
small as one atom diameter. This result implies that the 
thermodynamic stability of the macroscopic system does not  allow 
such a wide temperature width  $\delta T$ for fluctuation. 

The key to understanding this gradual fall of viscosity is the 
repulsive interaction.   
The particles with $\mbox{\boldmath $p$}\ne 0 $ in the flow are 
likely to behave similarly to other particles, 
especially with the particles having $\mbox{\boldmath $p$}=0$. 
If the former behave differently from the latter in the flow, the 
increase in the repulsive interaction energy is unavoidable. 
Hence, a considerable number of  particles show a singular dynamical behavior
even at the temperature for which not all particles participate in the BEC 
in the thermodynamical sense.
In such a case, although the BEC transition is a discontinuous change in 
the thermodynamic properties, it appears to be a continuous one in  the 
dynamical properties. If this picture is true, the kinematical approach 
will be useful for discussing $\nu (T)$ in the range of $T_{\lambda}<T<3.7$ K.
The microscopic formulation of the gradual fall of viscosity is a future problem.

\appendix

\section {Sum rule}
From the general principle, it is proved that the fluidity spectrum 
$\sigma(\omega)$ must satisfy the sum rule \cite {kub}. In  general, 
the susceptibility $\chi(\omega)$ is expressed by the response function $\phi(t)$ as 
\begin{equation}
  \chi(\omega)=\lim_{\epsilon\rightarrow 
  0}¥\int_{0}^{\infty¥}¥\phi(t)\exp(-i\omega t-\epsilon t)dt,
	   	\label{eq:3.5¥}
\end{equation}¥
while the relaxation function $\Phi(t)$ is expressed by this $\phi(t)$ as
\begin{equation}
  \Phi(t)=\lim_{\epsilon\rightarrow 
  0}¥\int_{t}^{\infty¥}¥\phi(t')\exp(-\epsilon t')dt'.
	   	\label{eq:3.55¥}
\end{equation}¥
The fluidity spectrum $\sigma(\omega)$ is expressed by this $\Phi(t)$ as
\begin{equation}
  \sigma(\omega)=\int_{0}^{\infty¥}¥\Phi(t)\exp(-i\omega t)dt.
	   	\label{eq:3.6¥}
\end{equation}¥
Hence, we obtain the rum rule
\begin{equation}
  \int_{-\infty¥}^{\infty¥}¥\sigma(\omega)d\omega=\pi\Phi(0),
	   	\label{eq:3.65¥}
\end{equation}¥
in which the right-hand side depends only on the basic parameters of 
the system. Such parameters determine the initial condition of relaxation at 
$t=0$, an example of which is $\pi\rho L/m¥$ in Eq.(\ref{eq:2978¥}).

\section {Chemical potential of the ideal Bose gas just above the 
transition temperature}

From the equation of states for $N/V$ in the ideal Bose, we obtain 
\begin{equation}
   1=\frac{1}{N¥}¥\frac{1}{e^{-\beta\mu}-1¥}¥
	 +  \frac{V}{N¥}¥\frac{g_{3/2}(e^{\beta\mu})}{\lambda^3¥}¥ ,
	\label{eq:3.651¥}
\end{equation}¥ 
where $g_{3/2}(z)=\sum_{n}^{\infty¥}¥z^{-n}/n^{3/2}$. 
The critical value $v_c$ at $T=T_0$ is given by
\begin{equation}
    v_c=\frac{\lambda_0^3¥}{g_{3/2}(1)}.
	\label{eq:3.652¥}
\end{equation}¥ 
Just above $T_0$, $V/N$ on the right-hand side of Eq.(\ref{eq:3.651¥}) can be 
approximated by  Eq.(\ref{eq:3.652¥}) as
\begin{equation}
    1=\frac{1}{N¥}¥\frac{1}{e^{-\beta\mu}-1¥}¥
	 + \left(\frac{\lambda _0¥}{\lambda}\right)^3¥¥\frac{g_{3/2}(e^{\beta\mu})}{g_{3/2}(1)¥}¥ .
	\label{eq:3.653¥}
\end{equation}¥ 
We are interested in $\mu(T)$ just above $T_0$, which is implicitly 
included in Eq.(\ref{eq:3.653¥}). For this purpose, the following  
expansion is useful \cite{robi}  
 \begin{eqnarray}
 g_{3/2}(e^{\beta\mu})=2.612
        &-&2\sqrt{\pi}¥(\beta\mu)^{0.5}+1.460\beta\mu-0.104(\beta\mu)^2
			    \nonumber \\
           &+& 0.00425(\beta\mu)^3-\cdots¥ .
	\label{eq:3.654¥}
\end{eqnarray}¥
 For $g_{3/2}(e^{\beta\mu})$ in Eq.(\ref{eq:3.653¥}), we use the first 
 and second terms in this expansion, and obtain Eq.(\ref{eq:3.653¥})
 in the first order of $\sqrt{\beta\mu}¥$ as
\begin{equation}
    1=\frac{1}{N|\beta\mu|}¥+\left(\frac{T}{T_0}\right)^{1.5}¥¥\left[1-\frac{2\sqrt{\pi}¥}{g_{3/2}(1)¥}¥\sqrt{|\beta\mu|}¥\right]¥ . 
	\label{eq:3.655¥}
\end{equation}¥ 
Defining $|\beta\mu|\equiv x$, we obtain
\begin{equation}
    \left(\frac{T}{T_0}\right)^{1.5}¥\frac{2\sqrt{\pi}¥}{g_{3/2}(1)¥}x^{1.5}
	                     +¥\left[1-¥\left(\frac{T}{T_0}\right)^{1.5}¥\right]¥ x=\frac{1}{N¥}¥.
	\label{eq:3.656¥}
\end{equation}¥ 
For a small $|\beta\mu|>1/N$, we can neglect $1/N$ on the right-hand 
side, and obtain
\begin{equation}
    x=\left(\frac{g_{3/2}(1)¥}{2\sqrt{\pi}¥}\right)^2¥
	                  ¥\left[\left(\frac{T_0}{T}\right)^{1.5}¥-1\right]^2¥ .
	\label{eq:3.657¥}
\end{equation}¥ 
Just above $T_0$, the right-hand side is  expanded in powers of $T-T_0$, hence
\begin{equation}
   	 \mu _0(T) = -\left(\frac{2.612\times 3}{4\sqrt{\pi}¥}¥\right)^2
	 k_BT_{0}\left(\frac{T-T_{0}}{T_{0}¥}\right)^{2} .
  	         \label{eq:3.858¥}
\end{equation}¥

\section {Combinatorial derivation of Eq.(\ref{eq:08¥})}
The partition function  $Z_0(N)$ of an ideal Bose gas is written
in a form reflecting the growth of the 
coherent wave function in configuration space, 
\begin{eqnarray}
	  \lefteqn{Z_0(N)=\frac{1}{N!¥}\left(\frac{m}{2\pi\beta\hbar^2¥}\right)^{3N/2} } \nonumber\\ 
	        &&  \times\int\sum_{per}¥  \exp\left[-\frac{m}{2\beta\hbar^2¥}
	            \sum_{i}^{N¥}(x_i-Px_i)^2¥¥¥\right] d^Nx_i¥¥,
				\label{eq:3.66¥}
\end{eqnarray}¥  
where $P$ denotes permutation.  
To calculate $Z_0(N)$, a geometrical consideration on the 
distribution of the coherent wave function is needed \cite {fey2}.

 (a)  Let us consider an elementary unit of the partition function $Z_0(N)$. 
 The $s$-size coherent wave function appears  
in the integral of Eq.(\ref{eq:3.66¥}) as
\begin{equation}
   ¥\int  \exp\left[-\frac{m}{2\beta\hbar^2¥}(x_{12}^2+\cdots+x_{s1}^2)¥¥¥\right]
             d^{s}x_i¥¥ \equiv L_s  ,
   \label{eq:3.67¥}
\end{equation}
where $x_{ij}^2=(x_i-x_j)^2$.  If the last term $x_{s1}^2$ in the exponent is replaced by $x_{s0}^2$, 
it becomes an open graph, and $L_s$ becomes a function of $x_1-x_0(\equiv x_{10})$ as  
\begin{equation}
 \widehat{L}_s(x_{10})=¥\int \exp\left[-\frac{m}{2\beta\hbar^2¥}(x_{12}^2+\cdots+x_{s0}^2)¥¥¥\right]
             d^{s-1}x_i¥¥ .
   \label{eq:3.68¥}
\end{equation}
This $\widehat{L}_s(x_{10})$ is related to $L_s$ as $L_s=\widehat{L}_s(x_{10}=0)$,
and is simply obtained  by $(s-1)$  
convolutions of $\exp (-mx_i^2/2\beta\hbar^2)$ with the initial one
$\exp (-mx_1^2/2\beta\hbar^2)$.  To obtain $\widehat{L}_s(x_{10})$, 
the convolution theorem states that one must calculate 
a three-dimensional Fourier-transformed function $\Gamma(p)$  for each $x$,
\begin{equation}
 \Gamma(p)=¥\int \exp\left[-\frac{mx^2}{2\beta\hbar^2¥}¥¥¥\right] \exp (ip\cdot x)d^{3}x¥¥ 
          =\lambda'^3\exp \left(-\frac{\lambda'^2p^2}{2¥}\right)¥,
   \label{eq:3.69¥}
\end{equation} 
where  $\lambda'=\lambda_t/\sqrt{2\pi}$ ($\lambda_t$ is the thermal wavelength).
Moreover, one must obtain an  inverse-Fourier-transformed function of $\Gamma(p)^s$ 
with respect to $x_{10}$ as 
 \begin{equation}
 \widehat{L}_s(x_{10}) = \frac{V}{(\sqrt{2\pi})^3¥}\int (\sqrt{2\pi})^{3(s-1)}\Gamma(p)^s
                 e^{-ipx_{10}}4\pi p^2dp¥,
   \label{eq:3.70¥}
\end{equation}
where $V$ comes from the translation of the closed graph.
Since $L_s=\widehat{L}_s(x_{10}=0)$, one has 
 \begin{equation}
 L_s= V\int (\sqrt{2\pi})^{3s}\Gamma(p)^s\frac{4\pi p^2dp}{(2\pi)^3¥}¥.
   \label{eq:3.71¥}
\end{equation}
 This integral, after using 
  \begin{equation}
 \int_{0}^{\infty¥}\exp (-ap^2)p^2dp=\frac{1}{8¥}¥\sqrt{\pi /a^3}¥,
   \label{eq:3.72¥}
\end{equation}
for  $\Gamma(p)$ in  Eq.(\ref{eq:3.69¥}), yields
\begin{equation}
 L_s= V\left(\lambda ^{3s}+ \frac{1}{2¥}¥\lambda ^{3(s-1)}\frac{V}{s^{3/2}¥}¥¥\right)¥,
    \label{eq:3.73¥}
\end{equation} 
where the first term $\lambda ^{3s}$ is obtained by extracting  $\Gamma(p=0)$ 
 from  Eq.(\ref{eq:3.71¥}).

(b) Let us consider a situation in which elementary closed graphs of size $s$ 
appear  $\xi_s$ times in $Z_0(N)$, with a distribution $\{\xi _1,\xi _2, 
\ldots,\xi_s,\ldots \}$ being subject to $N=\Sigma _{s}s\xi_s¥$. 
Consider the number of all possible 
configurations represented  by $\{\xi_1,\ldots,\xi_s,\ldots\}$, 
and denote it with $B(\xi_1,\ldots,\xi_s,\ldots)$. 
One can rewrite Eq.(\ref{eq:3.66¥}) as 
 \begin{eqnarray}
  Z_0(N)&=& \frac{1}{N!¥} \left(\frac{m}{2\pi\beta\hbar^2¥}\right)^{3N/2}   
			    \nonumber \\
  &\times& \sum_{\{\xi _s \}}B(\xi_1,\ldots,\xi_s,\ldots)L_1^{\xi_1}\cdots 
                 L_s^{\xi_s}\ldots¥¥.
				  \nonumber \\
    \label{eq:3.74¥}         
\end{eqnarray}¥
 $B(\xi_1,\ldots,\xi_s,\ldots)$ is estimated as follows.
Assume $N$ particles in an array.  The number of ways of partitioning them   
 into $\{\xi_1,\ldots,\xi_s,\ldots\}$ is given by $N!/\Pi _s \xi_s!$. 
 An array of $s$ particles corresponds to a closed graph of size $s$. 
For the coherent wave function, it does not matter which particle is 
the initial one in the array (circular permutation). Hence, $N!/\Pi _s \xi_s!$ 
must be multiplied by a factor $1/s$ for each $\xi_s$, with the result that
\begin{equation}
 B(\xi_1,\ldots,\xi_s,\ldots)=\frac{N!}{\prod_s \xi_s! s^{\xi_s}¥}¥.
    \label{eq:3.75¥}
\end{equation}
With this $B$ used in Eq.(\ref{eq:3.74¥}), one obtains
\begin{equation}
 Z_0(N)=\frac{1}{\lambda^{3N}}¥
        \sum_{\{\xi _s \}}\prod_s \frac{1}{\xi_s!¥}¥\left(\frac{L_s}{s}\right)^{\xi_s}¥.
    \label{eq:3.76¥}
\end{equation} 

 Using this $Z_0(N)$, the grand partition function 
\begin{equation}
   Z_0(\mu)=\Sigma _{N}¥Z_0(N)e^{\beta \mu N}
    \label{eq:3.765¥}
\end{equation}
 is obtained as follows.  The summation over $N$ under $\Sigma _{s}s\xi_s¥=N$
 changes to a free summation over $\xi_s$ from $0$ to $\infty $.  
 After substituting $N=\Sigma _{s}s\xi_s¥$ into $\lambda^{3N}$ 
in Eq.(\ref{eq:3.76¥}), and into $e^{\beta\mu N}$  in 
Eq.(\ref{eq:3.765¥}), this summation yields the exponential form
\begin{equation}
 Z_0(\mu)=\prod_s\exp \left[\frac{L_s}{s}
            \left(\frac{e^{\beta\mu}}{\lambda ^3¥}¥\right)^s¥\right]¥¥.
    \label{eq:3.77¥}
\end{equation}
Using Eq.(\ref{eq:3.73¥}) for $L_s$ in this $Z_0(\mu)$, one obtains 
Eq.(\ref{eq:08¥}).

\section {Spectral analysis of rotational superfluid-flow}

 The Hamiltonian in a coordinate system  rotating with a container is 
  $H-\mbox{\boldmath $\Omega$}\cdot \mbox{\boldmath $L$}$, where  
$\mbox{\boldmath $L$}$ is the total angular momentum and $\Omega$ is the 
applied angular  velocity.  The perturbation 
$H_{ex}=-\mbox{\boldmath $\Omega$}\cdot \mbox{\boldmath $L$}$ 
is cast in the form $-\sum_{i} (\mbox{\boldmath $\Omega$}\times 
\mbox{\boldmath $r$})\cdot \mbox{\boldmath $p$}¥$ , in which  
$\mbox{\boldmath $\Omega$}\times \mbox{\boldmath $r$} \equiv \mbox{\boldmath $v$}_d( \mbox{\boldmath $r$})$ 
serves as the external field. 
Defining a mass-current density $\mbox{\boldmath $J$}(\mbox{\boldmath 
$r$})$, we express  the perturbation $H_{ex}$ as
\begin{equation}
   -\mbox{\boldmath $\Omega$}\cdot \mbox{\boldmath $L$}
         =-\int  \mbox{\boldmath $v$}_d( \mbox{\boldmath $r$})\cdot 
   \mbox{\boldmath $J$}(\mbox{\boldmath $r$}) d^3x.¥
               \label{eq:3.781¥}
\end{equation}¥
Because of $ div\mbox{\boldmath $v$}_d( \mbox{\boldmath $r$})=0 $, 
 $\mbox{\boldmath $v$}_d( \mbox{\boldmath $r$})$
 acts as a transverse-vector probe to the excitation of bosons.
 For the rotation, one must use the transverse susceptibility $\chi ^T (q,\omega)$ for 
$ \mbox{\boldmath $v$}_d( \mbox{\boldmath $r$} )$ such as 
$\mbox{\boldmath $J$} (\mbox{\boldmath $r$})=\left[\lim_{q\rightarrow 0}
 \chi ^T (q,0)\right] \mbox{\boldmath $v$}_d( \mbox{\boldmath $r$} )$. 
Using $ \mbox{\boldmath $\Omega$} =(0,0,\Omega) $ on the left-hand 
side of Eq.(\ref{eq:3.781¥}), 
and using the above $\mbox{\boldmath $J$} (\mbox{\boldmath $r$})$ and 
$\mbox{\boldmath $v$}_d=\mbox{\boldmath $\Omega$}\times \mbox{\boldmath $r$}=(-\Omega y, \Omega x,0) $ 
on its right-hand side, one obtains the angular momentum $L_z$ as
\begin{equation}
     L_z=  \chi^T(0,0)\int_{V}(x^2+y^2) d^3x \cdot\Omega .¥
	              \label{eq:3.782¥}
\end{equation}¥
In a normal fluid, the susceptibility satisfies $\chi^T(0,0)=\chi 
^L(0,0)$, and therefore the ordinary use of $\rho$ is justified. 
The classical moment of inertia is given by
\begin{equation}
     I_z^{cl}=  mn\int_{V}(x^2+y^2) d^3x = \chi^L(0,0)\int_{V}(x^2+y^2) d^3x .¥
	         \label{eq:3.783¥}
\end{equation}¥
In a superfluid,  however, we must go back to Eq.(\ref{eq:3.782¥}), and 
 write the moment of inertia $I_z=L_z/\Omega$  as
\begin{equation}
      I_z=I_z^{cl}\left(1- \frac{1}{\rho¥}¥\lim_{q\rightarrow 0} 
                      [\chi^L(q,0)-\chi^T(q,0)¥]\right)¥.
					   \label{eq:3.784¥}
\end{equation}¥
For the dynamic response, the current $\mbox{\boldmath $J$}(\mbox{\boldmath $r$})=\chi^T(0,0)
\mbox{\boldmath $v$}_d( \mbox{\boldmath $r$})$ is replaced by
 $\mbox{\boldmath $J$}(\mbox{\boldmath $r$})=\chi^T(0,0,\Omega, r)
\mbox{\boldmath $v$}_d( \mbox{\boldmath $r$})$, where 
 $r$=$\sqrt{x^2+y^2}¥$ is the distance from the center of rotation. 
Correspondingly,  we define
\begin{equation}
     L_z= \int_{V}\chi^T(0,0, \Omega, r)r^2 d^3x \cdot \Omega ,¥
	            \label{eq:3.785¥}
\end{equation}¥
and
\begin{equation}
     I_z(\Omega)= I_z^{cl} - \lim_{q\rightarrow 0} \int_{V}
                       [\chi^L(q,0,\Omega, r)-\chi^T(q,0,\Omega, r)¥]
                         r^2 d^3x .¥
						     \label{eq:3.786¥}
\end{equation}¥
The position-dependent angular velocity is defined as
\begin{equation}
     \mbox{\boldmath $\Omega$}_0(\mbox{\boldmath $r$})= 
                   \left(1-\frac{1}{\rho¥}¥\lim_{q\rightarrow 0} 
                    [\chi^L(q,0,\Omega, r)-\chi^T(q,0,\Omega, r)¥]\right)¥¥\mbox{\boldmath $\Omega$}.
					        \label{eq:3.787¥}
\end{equation}¥
  Let us consider the first approximation of the above quantities.  We begin  with 
   \begin{equation}
   	 <J_{\mu}(x,t)>= <G|S^{\dagger}\hat{J}_{\mu}(x,t)S|G>,
	            \label{eq:3.788¥}
   \end{equation}¥
where $S= T exp\left[-i\int_{-\infty}^{t}dt' \hat{H}_{ex}(\mbox{\boldmath $r$},t')¥\right]¥$.
  Using $H_{ex}(\mbox{\boldmath $r$})=-v^{\mu}_{d}(\mbox{\boldmath 
  $r$})J_{\mu}(\mbox{\boldmath $r$})$, the analytical continuation 
  $t\rightarrow \tau =it$ is performed in the higher-order expansion terms 
on the right-hand side of Eq.(\ref{eq:3.788¥}). 
 As the simplest nonlinear susceptibility for $J_{\mu}$, we consider the third-order term 
 $\chi^{(3)}_{\mu,\nu\sigma\tau}v_d^{\nu}v_d^{\sigma}v_d^{\tau}$ with 
 respect to $H_{ex}$, and extract  a correction term to the linear  
 susceptibility  $\chi^{(3)}_{\mu\nu}(v_d)\mbox{\boldmath $v$}_d$ from  
 $\chi^{(3)}_{\mu,\nu\sigma\tau}v_d^{\nu}v_d^{\sigma}v_d^{\tau}$ as
  \begin{eqnarray}
     \chi^{(3)}_{\mu\nu}(q,i\omega)
     &=& \beta n_0|\mbox{\boldmath $v$}_d( \mbox{\boldmath $r$})|^2¥
       \frac{1}{\beta ^2¥¥}¥\sum_{n,m}¥\frac{1}{V^2¥}\sum_{p,p_1}\left(p+\frac{q}{2¥}¥\right)_{\mu}
	                     \nonumber\\ 
	  & \times & \left(p_1+\frac{q}{2¥}¥\right)_{\nu}
                  \left(\frac{\mbox{\boldmath $p$}+\mbox{\boldmath $p_1$}}{2¥}¥\right)\cdot 
                  \left(\frac{\mbox{\boldmath $p$}+\mbox{\boldmath $p_1$}}{2¥}+\mbox{\boldmath $q$}¥\right)  
				      \nonumber\\ 
     & \times & G(i\omega _n+i\omega,p+q) G(i\omega _n,p)
	                   \nonumber\\ 
	 & \times & G(i\omega _m+i\omega,p_1+q) G(i\omega _m,p_1),
	                 \nonumber\\ 
	                \label{eq:3.789¥}
\end{eqnarray}¥
where
\begin{equation}
	 G(i\omega _n,p)=\frac{1}{i\omega _n - \epsilon (p)-\Sigma +\mu¥}¥.
	             \label{eq:3.7901¥}
\end{equation}¥
As a result, we obtain
\begin{equation}
	\hat{\chi}^{(3)}_{\mu\nu}(q,0,\Omega,r)= -q_{\mu}q_{\nu}\frac{1}{V^2¥}
	               \left(\frac{q^2}{4}\right)¥ 
	               |F_{\beta}(q)|^2 (\Omega r)^2\beta n_0(T),
	             \label{eq:3.791¥}
\end{equation}¥
where
\begin{equation}
	 F_{\beta}(q)= \frac{(\exp(\beta[\Sigma-\mu]-1))^{-1}-(\exp(\beta[\epsilon (q)+\Sigma-\mu])-1)^{-1}} 
	                         {\epsilon (q)¥¥} ¥,
	             \label{eq:3.792¥}
\end{equation}¥
 is a positive monotonically decreasing function of $q^2$, which approaches zero as 
$q^2\rightarrow \infty$. The infinite sum of the bubble-chain 
diagrams due to the particle interaction is given by 
 \begin{eqnarray}
	\hat{\chi}^{(3)}_{\mu\nu}(q,0,\Omega,r)
	      &=& -\frac{q_{\mu}q_{\nu}}{2}\left(\frac{q^2}{2}\right)\frac{1}{V^2¥}\frac{1}{n^2¥}
	                        \frac{|F_{\beta}(q)|^2}{\left[1-UF_{\beta}(q)\right]^2¥}(\Omega r)^2
							 \nonumber\\ 
				 & \times &	\beta n_0(T)¥.
	     \label{eq:3.793¥}
\end{eqnarray}¥
Hence, we obtain
 \begin{eqnarray}
	\hat{\chi}^{(3)}_{\mu\nu}(q,0,\Omega,r)
	              &=& -\frac{1}{V^2¥}\left(\frac{2m}{U\beta_{on}}\right)^2 \beta_{on}\frac{1}{n^2¥}¥ 
	                           \nonumber\\ 
            & \times &  \tanh ^2 \left(\frac{\beta_{on}[\mu(T_{on})-\Sigma]}{2¥}¥\right)¥ (\Omega r)^2
			                     \nonumber\\ 
			 & \times &    n_0(T_{on}) \frac{q_{\mu}q_{\nu}}{q^2}¥.
						   \nonumber\\ 
                  \label{eq:3.794¥}
\end{eqnarray}¥
Using this result in the next-order term in Eq.(\ref{eq:3.787¥}), we 
obtain the $\Omega$-dependent form of the angular velocity of a superfluid 
in a rotating basket \cite {busk} as
\begin{equation}
     \mbox{\boldmath $\Omega$}_0(\mbox{\boldmath $r$})= 
                   \left[1-\frac{\hat{\rho _s}(T)}{\rho¥}¥+\frac{1}{c(T)¥}¥¥\left(\frac{\hat{\rho _s}(T)}{\rho¥}¥\right)^3
                         \frac{m(\Omega r)^2}{k_BT¥}¥\right]¥
                           ¥¥\mbox{\boldmath $\Omega$}.
						           \label{eq:3.795¥}
\end{equation}¥

¥


\begin{thebibliography}{99}
\bibitem{inv}  Bifurcation is a phenomenon in which a simple dynamical process 
 qualitatively  changes to a more complex one when an
 external parameter exceeds a critical value, such as the transition of 
a laminar flow to a turbulent flow.
\bibitem{tur}  The oscillatory Poiseuille's flow is often used in the 
experiments on turbulence. The flow velocity $v$ and angular frequency 
$\omega$  in our thought experiment on a superfluid flow
are much smaller than those used in the experiments on turbulence.
\bibitem{axis}  Whereas the following argument is applicable to any $r$, we 
consider  $\mbox{\boldmath $v$}$ along the axis of the capillary ($r=0$) for simplicity.
\bibitem{unit} $h(s)$ is a quantity per unit volume as in  Eq.(\ref{eq:09¥}).
\bibitem{dam} The propagation of transverse oscillations in 
a liquid is strongly damped, but the second term in the bracket of 
Eq.(\ref{eq:2995¥}) represents the transverse 
one-particle excitations from the  
condensate, being a phenomenon before the damping.
\bibitem{pre} The capillary method is not suitable for the precise measurement of 
 $\nu(T)$, which is normally performed by the oscillating-disc viscometer or the rotation 
viscometer. In contrast to the $\nu(T)$ by a capillary, however,  the $\nu(T)$ by the viscometer
does not show an abrupt drop, but a continuous fall. The range of $2.17K<T<2.18K$ 
is the region in which such a continuous $\nu(T)$ shows a steepest  
 fall \cite {don}. We cannot more precisely determine the temperature 
 at which $\nu(T)$ abruptly drops in the capillary experiment.
\end{thebibliography}
\end{document}